\documentclass[runningheads]{cl2emult}

\usepackage{makeidx}  % allows index generation
\usepackage{graphicx} % standard LaTeX graphics tool
\usepackage{epsfig}
\usepackage{subeqnar} % subnumbers individual equations
\usepackage{multicol} % used for the two-column index
\usepackage{cropmark} % cropmarks for pages without
\usepackage{lnp}      % placeholder for figures
\usepackage{amsmath}
\usepackage{amssymb}
\usepackage{latexsym}
\makeindex            % used for the subject index
\begin{document}
\title*{Free cooling of particles with rotational degrees of freedom}
\toctitle{Free cooling of particles with rotational degrees of freedom}
% allows explicit linebreak for the table of content
%
%
\titlerunning{Free cooling of particles}
% allows abbreviation of title, if the full title is too long
% to fit in the running head
%
\author{Timo Aspelmeier
\and Martin Huthmann
\and Annette Zippelius}
\authorrunning{Timo Aspelmeier et al. }
% if there are more than two authors,
% please abbreviate author list for running head
%
%
\institute{Institut f\"ur Theoretische Physik, Universit\"at
  G\"ottingen,\\ D-37073 G\"ottingen, Germany}

\maketitle              % typesets the title of the contribution

\begin{abstract} 
  Free cooling of granular materials is analyzed on the basis of a
  pseudo-Liouville operator. Exchange of translational and rotational
  energy requires surface roughness for spherical grains, but occurs
  for non-spherical grains, like needles, even if they are perfectly
  smooth. Based on the assumption of a homogeneous cooling state, we
  derive an approximate analytical theory. It predicts that cooling of
  both rough spheres and smooth needles proceeds in two stages: An
  exponentially fast decay to a state with stationary ratio of
  translational and rotational energy and a subsequent algebraic decay
  of the total energy. These results are confirmed by simulations for
  large systems of moderate density. For higher densities, we observe
  deviations from the homogeneous state as well as large-scale
  structures in the velocity field. We study non-Gaussian
  distributions of the momenta perturbatively and observe a breakdown
  of the expansion for particular values of surface roughness and
  normal restitution.
\end{abstract}

\section{Introduction}
\label{Introduction}
The hard-sphere model has been a very useful reference system for our
understanding of classical liquids \cite{Hansen}. As far as static
correlations are concerned, an analytical expression for the pair
correlation is available \cite{Thiele63} and provides a good first
approximation for particles interacting via smooth-potential
functions. The hard-sphere model is even more important for the
dynamics, because it allows for approximate analytical solutions,
based on the Boltzmann equation and its generalization by Enskog to
account for a finite particle diameter and pair correlations at
contact \cite{Lebowitz69}. The model has the additional advantage that
it is particularly well suited for numerical simulations
\cite{Alder70} and in fact many of the important phenomena of dense
liquids have been observed first in simulations of hard spheres.
Examples are the discovery of long-time tails \cite{Erpenbeck81} and
two-dimensional solids \cite{Alder62}.

Not surprisingly the model has become very popular also in the context
of granular media, which are characterized by inelastic collisions of
their constituents. Focusing on the rapid-flow regime, where kinetic
theory should apply, generalized Boltzmann- and Enskog equations have
been formulated and solved approximately
\cite{general,jenkins85b,lun87,lun91,Goldshtein95}. The success of the
Boltzmann-Enskog equation in classical fluids is based on the
linearisation of the collision operator around local equilibrium.  The
resulting linear hermitean operator can then be treated by standard
methods of functional analysis \cite{Grad,Waldmann,chapman60}. For
inelastic systems no analog of the local equilibrium distribution is
known. In many studies, including the present one, a homogeneity
assumption is made, which is known to be unstable for dense and large
enough system and long times \cite{goldhirsch93}.  Hence the analysis
is restricted to small and intermediate densities.  Alternatively, one
may restrict oneself to almost elastic collisions and expand around
the elastic case.

Kinetic theory of rough, inelastic, circular disks was first discussed
by Jenkins and Richman \cite{jenkins85b}. These authors introduced two
temperatures, one for the translational and one for the rotational
degrees of freedom, and studied deviations from a two-temperature
Maxwellian distribution, using Grad's moment expansion.  Subsequently
Lun and Savage \cite{lun87,lun91} extended the approach to rough,
inelastic spheres. A set of conservation equations and constitutive
relations was derived from the Boltzmann equation, assuming small
inelasticity and surface roughness. Goldshtein and Shapiro
\cite{Goldshtein95} discuss in detail the homogeneous cooling state of
rough spheres. They determine the asymptotic ratio of rotational to
translational energy as a function of surface roughness and
coefficient of normal restitution.  Hydrodynamic equations and
constitutive relations are derived with help of the Enskog expansion.
More recently, event-driven simulations of rough spheres have been
performed by McNamara and Luding \cite{mcnamara98}. They investigate
free cooling as a function of arbitrary surface roughness and normal
restitution and compare their results to an approximate kinetic theory
\cite{huthmann97,luding98huth}.

Most analytical and numerical studies of kinetic phenomena have
concentrated on spherical objects so far\footnote{Exceptions are
  computer simulations of polygonal particles \cite{polygons} and
  cellular automata models \cite{cellular}}. The question then arises,
which of the results are specific to spherical objects and which are
generic for inelastically colliding particles. A single collision of
two arbitrarily shaped, but convex objects is quite difficult to
describe analytically \cite{brogliato96}, set aside the problem of an
ensemble of colliding grains. In this paper we have chosen the
simplest non spherical objects, needles, which allow for an
analytical, albeit approximate solution and large scale simulations
\cite{huthasp}.

The paper is organized as follows. In Sec.\ref{Section2} we introduce
the time evolution operator. For pedagogical reasons we first discuss
smooth potentials and recall the formalism of a pseudo-Liouville
operator for elastic, hard-core collisions. Subsequently the formalism
is extended to inelastic, rough spheres and needles. The homogeneous
cooling state is introduced in Sec. \ref{HCS}. We present results for
both spheres and needles, assuming a Maxwellian distribution for
linear and angular momenta. We show with simulations that for dense
systems of needles the assumption of homogeneity breaks down.
Corrections to a Gaussian approximation are discussed in Sec.
\ref{Expansion}. Finally in Sec. \ref{Conclusions} we summarize
results and present conclusions. Some details of the calculation are
delegated to appendices.

\section{The Liouville operator}
\label{Section2}
We are interested in macroscopic properties of systems of many particles which
are themselves meso- or macroscopic, i.e. behave according to the laws
of classical
mechanics as opposed to quantum mechanics. In addition, our systems are
\textit{granular} so energy is not conserved. This means that they can not be
treated  with Hamiltonian mechanics. We will present here a formalism
based on the Liouville operator that enables us nevertheless to derive
properties of the system under consideration.

We consider two different models: The first is a system of spheres of
diameter $d$ and the second is one of (infinitely) thin rods or
needles of length $L$. In order to keep the discussion as transparent
as possible, the formalism of the (pseudo) Liouville operator will be
demonstrated for Hamiltonian systems with smooth potentials first, for
hard core potentials next, and finally for granular spheres and
needles.  It is interesting to note that both cases, spheres and
needles, are analytically tractable so that comparisons between
different geometrical particle shapes are possible.

\subsection{Smooth potentials}
We consider a system of $N$ classical particles of mass $m$ in a
volume $V$, interacting through pairwise potentials. The system is
characterized by its total energy
\begin{equation}
H=\sum_{i=1}^N \frac{\vec{p}_i^2}{2m}+\sum_{i<j}U(\vec{r}_i-\vec{r}_j)
\end{equation}
in terms of particle momenta $\vec{p}_i$ and coordinates $\vec{r}_i$. The time
evolution of an observable $f(\Gamma)$, which is a function of phase space
variables $\Gamma:=\{\vec{r}_i,\vec{p}_i\}$, but does not depend on time
explicitly, is given in terms of the Poisson bracket by
\begin{equation}
\label{eq:lioudef}
\frac{df}{dt} = \{H, f\}  =: i\mathcal{L}f.
\end{equation}
This defines the Liouville operator $\mathcal{L}$. The time evolution
of $f$ can then formally be written in terms of $\mathcal{L}$: 
$f(t) = e^{i\mathcal{L}t}f(0)$.

We decompose the Liouville operator $\mathcal{L}= \mathcal{L}_0+
\mathcal{L}_{\text{inter}}$ into a free-streaming part $\mathcal{L}_0$
and an operator $\mathcal{L}_{\text{inter}}$, which accounts for
interactions. The definition of the Poisson bracket,
\begin{equation}
\{H,f\}=\sum_j\left(\frac{\partial f}{\partial \vec{r}_j}
\frac{\partial H}{\partial \vec{p}_j}-
\frac{\partial f}{\partial \vec{p}_j}
\frac{\partial H}{\partial \vec{r}_j}\right),
\end{equation}
thus yields
\begin{align}
\label{eq:freestream}
i \mathcal{L}_0 &= \sum_{j}
  \frac{\vec{p}_j}{m}\frac{\partial}{\partial \vec{r}_j}\quad\text{and}\quad
i \mathcal{L}_{\text{inter}} = \sum_{k<j} 
\frac{\partial U}{\partial \vec{r}_{kj}}
\left(\frac{\partial}{\partial \vec{p}_j}-
\frac{\partial}{\partial \vec{p}_k}\right)
\end{align}
with $\vec{r}_{ij}=\vec{r}_i-\vec{r}_j$.

\subsection{Elastic hard-core interactions}
A pseudo-Liouville operator for hard-core collisions has been
formulated by Ernst et al.  \cite{ernst69} and has been applied by
many groups \cite{Resibois} to study the dynamic evolution of a gas of
hard spheres. Collisions are instantaneous and characterized by
collision rules. In a collision of two particles, numbered $1$ and
$2$, their pre-collisional velocities $\vec{v}_1=\vec{p}_1/m$ and
$\vec{v}_2=\vec{p}_2/m$ are changed instantaneously to their
post-collisional values $\vec{v}'_1$ and $\vec{v}'_2$ according to
\begin{equation}
\label{eq:postvelel}
\begin{split}
\vec{v}'_1&=\vec{v}_1-(\vec{v}_{12}\hat{\vec{r}}_{12})\hat{\vec{r}}_{12}
\\
\vec{v}'_2&=\vec{v}_2+(\vec{v}_{12}\hat{\vec{r}}_{12})\hat{\vec{r}}_{12}.
\end{split}
\end{equation}
We have denoted the relative velocity by
$\vec{v}_{12}=\vec{v}_{1}-\vec{v}_{2}$, and $\hat{\vec{r}}=\vec{r}/|\vec{r}|$.
The free-streaming part of the Liouville operator remains unchanged, whereas
the part which accounts for interactions has to be modified because the
potential is no longer differentiable in the limit of hard-core interactions.
As a consequence, $\mathcal{L}$ is no longer self adjoint as it is for systems
with smooth potentials.  This is why it is called a pseudo-Liouville operator
for hard-core systems. For the same reason we will need \textit{two} Liouville
operators below, one for forward and one for backward time evolution.

In order to construct the pseudo Liouville operator, we consider the
change of a dynamical variable due to a collision of just two
particles. What we need is an operator $\mathcal{T}_+^{(12)}$ that
\begin{itemize}
\item gives the change of an observable through a collision when
  integrated over a short time interval containing the collision time (since
  the hard core interaction is non-differentiable, we have to resort
  to integrating over the collision instead of looking at the
  derivatives directly),
\item only acts at the time of contact,
\item only acts when the particles are approaching but not when they
  are receding.
\end{itemize}
The second requirement can be satisfied by $\mathcal{T}_+^{(12)}
\propto \delta(|\vec{r}_{12}|-d)$, the third one demands
$\mathcal{T}_+^{(12)} \propto \Theta(-\frac{d}{dt}|\vec{r}_{12}|)$,
where $\Theta(\cdot)$ is the usual Heaviside step function.  In order
to satisfy the first point, we use an operator $b_+^{(12)}$ which is
defined by its action on an observable $f$ according to
\begin{equation}
b_+^{(12)} f(\vec{v}_1,\vec{v}_2) =
f(\vec{v}_1',\vec{v}_2'),
\end{equation}
i.e. it simply replaces all velocities according to eqs.
\eqref{eq:postvelel}. The operator $\mathcal{T}_+^{(12)}$ should give
the \textit{change} induced by a collision, so that
$\mathcal{T}_+^{(12)} \propto b_+^{(12)} - 1$.  We collect the three
terms  and make sure to include a prefactor which is chosen
such that the integration of an observable over a short time interval
around the collision time yields the change of the observable, as
induced by the collision rules \eqref{eq:postvelel}.  The complete
expression for $\mathcal{T}_+^{(12)}$ is thus
\begin{equation}
\label{eq:lioucoll}
i \mathcal{T}_+^{(12)} = \left|\frac{d}{dt}|\vec{r}_{12}|\right|
                \delta(|\vec{r}_{12}|-d)
                \Theta(-\frac{d}{dt}|\vec{r}_{12}|)
                (b_+^{(12)} - 1).
\end{equation}
Since the probability that three or more particles touch at precisely
the same instant is zero, we only need to consider two-particle
collisions and find for the time-evolution operator for the system of
elastically colliding hard spheres:
\begin{align}
f(t)&=e^{i(\mathcal{L}_0+\mathcal{L}_{\pm})t}f(0) \quad\text{ for }
   t\gtrless 0
\intertext{with}
\label{Liouville}
i \mathcal{L}_{\pm} &= \sum_{i<j}i\mathcal{T}_\pm^{(ij)} 
              = \sum_{i<j}
                \left|\frac{d}{dt}|\vec{r}_{ji}|\right|
                \delta(|\vec{r}_{ji}|-d)
                \Theta\left(\mp\frac{d}{dt}|\vec{r}_{ji}|\right)
                (b_\pm^{(ij)} - 1).
\end{align}
The negative time evolution is given by $\mathcal{L}_-$, and
$b_-^{(ij)}$ is the operator that replaces post-collisional velocities
by pre-collisional ones.

\subsubsection*{Extension to rough spheres}

Hard-core models of elastically colliding spheres have been extended
to include {\it rotational} degrees of freedom and {\it surface
  roughness} \cite{jenkins85b,chapman60}. Rotational degrees of
freedom offer the possibility to describe molecules with internal
degrees of freedom and surface roughness is needed to transfer energy
from the translational degrees of freedom to the rotational ones.

We only discuss the simplest case of identical spheres of mass $m$,
moment of inertia $I$ and diameter $d$. Translational motion is
characterized by the center-of-mass velocities $\vec{v}_i$ and
rotational motion by the angular velocities $\vec{\omega}_i$.   Let the surface normal $\hat{\vec{r}}_{12}$ at the point of
contact point from sphere 2 to sphere 1. The important quantity to
model the collision is the relative velocity of the point of contact:
\begin{equation}
\vec{V} = (\vec{v}_1 - \frac{d}{2}\vec{\omega}_1\times\hat{\vec{r}}_{12}) -
             (\vec{v}_2 + \frac{d}{2}\vec{\omega}_2\times\hat{\vec{r}}_{12}).
\end{equation}
There are two contributions, firstly the center of mass velocity of
each sphere, and secondly the contributions from the rotations of each
sphere.  The minus sign in the first parenthesis stems from the fact
that the surface normal, as it was defined, points outwards for sphere
2 and inwards for sphere 1.

Now we can specify the collision rules. Primed variables always denote
quantities immediately after the collision; unprimed variables denote
pre-collisional quantities:
\begin{equation}
\label{eq:rulenormalsphere}
\begin{split}
\hat{\vec{r}}_{12}\vec{V}' &= - \hat{\vec{r}}_{12}\vec{V} \\
\hat{\vec{r}}_{12}\times\vec{V}' &= -e_t\, \hat{\vec{r}}_{12}\times\vec{V}.
\end{split}
\end{equation}

As we are still dealing with elastic spheres, energy conservation
requires $e_t=+1$, corresponding to perfectly rough spheres, where the
tangential velocity component is completely reversed. Perfectly smooth
spheres $e_t=-1$ are also compatible with energy conservation, but
reduce to the above simple case of spheres without rotational degrees
of freedom, because during collision the angular velocities remain
unchanged. Later, we will also admit other values for $e_t$.

Eqs. \eqref{eq:rulenormalsphere} form three linearly independent
equations. In addition, total momentum is conserved,
\begin{equation}
\label{eq:linmomcons}
  \vec{v}_1' + \vec{v}_2' = \vec{v}_1 + \vec{v}_2,
\end{equation}
and forces during a collision can only act at the point of contact.
Therefore there is no torque with respect to this point and
consequently we have conserved angular momentum (also with respect to
the point of contact) for \textit{both} particles involved:
\begin{equation}
\label{eq:angmomcons}
\begin{split}
\frac{md}{2}\hat{\vec{r}}_{12}\times(\vec{v}_1'-\vec{v}_1) + 
I(\vec{\omega}_1'-\vec{\omega}_1) = \vec{0} \\
\frac{md}{2}\hat{\vec{r}}_{12}\times(\vec{v}_2'-\vec{v}_2) - 
I(\vec{\omega}_2'-\vec{\omega}_2) = \vec{0}.
\end{split}
\end{equation}
Altogether we have 12 independent equations for 12 unknowns, namely
the four vectors $\vec{v}_i'$ and $\vec{\omega}_i'$ with three
components each. Solving for these, we obtain:
\begin{equation}
\label{eq:postvelsphere}
\begin{split}
\vec{v}_1' &= \vec{v}_1 - \eta_t\vec{v}_{12} - 
                 (\eta_n-\eta_t)(\hat{\vec{r}}_{12}\vec{v}_{12})\hat{\vec{r}}_{12} -
                 \eta_t\frac{d}{2}\hat{\vec{r}}_{12}\times
                 (\vec{\omega}_1+\vec{\omega}_2) \\ 
\vec{v}_2' &= \vec{v}_2 + \eta_t\vec{v}_{12} + 
                 (\eta_n-\eta_t)(\hat{\vec{r}}_{12}\vec{v}_{12})\hat{\vec{r}}_{12} +
                 \eta_t\frac{d}{2}\hat{\vec{r}}_{12}\times
                 (\vec{\omega}_1+\vec{\omega}_2) \\
\vec{\omega}_1' &= \vec{\omega}_1 + \frac{2}{dq}\eta_t\hat{\vec{r}}_{12}\times
                      \vec{v}_{12} +
                      \frac{\eta_t}{q}\hat{\vec{r}}_{12}\times(\hat{\vec{r}}_{12}\times
                      (\vec{\omega}_1+\vec{\omega}_2)) \\
\vec{\omega}_2' &= \vec{\omega}_2 + \frac{2}{dq}\eta_t\hat{\vec{r}}_{12}\times
                      \vec{v}_{12} +
                      \frac{\eta_t}{q}\hat{\vec{r}}_{12}\times(\hat{\vec{r}}_{12}\times
                      (\vec{\omega}_1+\vec{\omega}_2)).
\end{split}
\end{equation}
The dimensionless constant $q = 4I/(md^2)$ abbreviates a frequently
appearing combination of factors.  We have also introduced two
parameters $\eta_n$ and $\eta_t$, because we anticipate the more
general collision rules for the inelastic case.  For elastically
colliding spheres, we simply have $\eta_n=1$ and $\eta_t=q/(1+q)$ for
perfectly rough and $\eta_t=0$ for perfectly smooth spheres.

The pseudo-Liouville operator for elastically colliding rough spheres
is still given by eq. \eqref{Liouville} but the operator $b_+^{(ij)}$
now replaces linear \textit{and} angular velocities according to
eqs.  \eqref{eq:postvelsphere}.

\subsubsection*{Extension to rough needles}
Elastic collisions of hard needles have been discussed by Frenkel et
al. \cite{frenkel}. It is straightforward to rephrase their results in
terms of a pseudo-Liouville operator \cite{huthasp}. The free
streaming part of the Liouville operator is derived from the kinetic
energy of the Hamiltonian according to the general rules of classical
dynamics. Note however, that for  needles, one of the moments of
inertia is zero; this implies that the angular-momentum component
along the corresponding axis, which points along the orientation of
the needle, is also always zero. Therefore, rotations about this axis
can be ignored, and $\vec{\omega}$ has only two components, both
perpendicular to the orientation of the needle.  The center of mass
coordinate of needle $i$ will be denoted by $\vec{r}_i$ and its
orientation by the unit vector $\vec{u}_i$. The moments of inertia
perpendicular to $\vec{u}_i$ are equal due to symmetry and will be
denoted by $I$.

\begin{figure}[hbt]
\centering
\includegraphics[width=.4\textwidth]{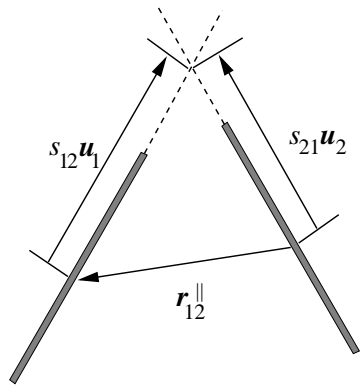}
\caption[]{Configuration of two needles projected in the plane spanned
  by the unit vectors $\vec{u}_1$ and $\vec{u}_2$
}
\label{fig:lioucoll1}
\end{figure}

The formulation of the collision rules proceeds in close analogy to
rough spheres. First we determine the conditions of contact.  The unit
vectors $\vec{u}_1$ and $\vec{u}_2$ span a plane $E_{12}$ with normal
\begin{equation}
\vec{u}_\perp =
  \frac{\vec{u}_1 \times \vec{u}_2} {|\vec{u}_1 \times
  \vec{u}_2|}. \label{usenk}
\end{equation}
We decompose $\vec{r}_{12}=\vec{r}_1-\vec{r}_2$ into a component
perpendicular $\vec{r}_{12}^\perp =
(\vec{r}_{12}\vec{u}_\perp)\vec{u}_\perp$ and parallel
$\vec{r}_{12}^{\parallel} = (s_{12}\vec{u}_1-s_{21}\vec{u}_2) $ to
$E_{12}$ (see fig. \ref{fig:lioucoll1}).  The rods are in contact if
%$|\vec{r}_{12}^\perp|=0^+$ (we take the proper limit of infinitly thin
%rods) 
$\vec{r}_{12}\vec{u}^\perp = 0 $ 
and simultaneously $|s_{12}|<L/2$ and $|s_{21}|<L/2$.

The relative velocity of the point of contact is given by
\begin{equation}\label{vrel}
\vec{V} = \vec{v}_{12} + s_{12}\dot{\vec{u}}_1 - s_{21}\dot{\vec{u}}_2.
\end{equation}
It is useful to introduce a set of normalized basis vectors
\begin{equation}
  \label{orthosys}
\vec{u}_1,\quad \vec{u}_1^\perp=
(\vec{u}_2-(\vec{u}_1\vec{u}_2)\vec{u}_1)/\sqrt{1-(\vec{u}_1\vec{u}_2)^2}
,\quad {\rm 
and}  \quad \vec{u}_{\perp}
\end{equation}
with $\vec{u}_{\perp}$ defined in eq. (\ref{usenk}).  Total momentum
conservation is given by \eqref{eq:linmomcons} and conservation of
angular momentum with respect to the contact point reads
\begin{equation}
  \label{drstimp}
 \vec{\omega}_1^{'}=\vec{\omega}_1
  +\frac{m s_{12}}{I}\vec{u}_1\times  (\vec{v}_1'- \vec{v}_1)\quad
  {\rm and}\quad
  \vec{\omega}_2^{'}=
  \vec{\omega}_2+\frac{m s_{21}}{I}\vec{u}_2\times (\vec{v}_2'- \vec{v}_2).
\end{equation}
Three additional equations follow from the change in the
relative velocity of the contact point, which is modeled in close
analogy to the case of rough spheres:
\begin{equation}
\vec{V}'\vec{u}_\perp = -\vec{V}\vec{u}_\perp, \quad 
\vec{V}'\vec{u}_1 = -e_t\vec{V}\vec{u}_1, \quad {\rm and} \quad 
\vec{V}'\vec{u}_2 = -e_t\vec{V}\vec{u}_2.
\end{equation}
Again, energy conservation implies $e_t=\pm 1$, corresponding to either
perfectly rough or perfectly smooth needles (see also eq. (\ref{needloss})).
Solving for $\vec{v}_i'$ and $\vec{\omega}_i'$, we obtain after a lengthy
calculation:
\begin{equation}
\label{eq:postvelneedle}
\vec{v}_1' = \vec{v}_1+\Delta\vec{v} \quad {\rm and} \quad 
\vec{v}_2' = \vec{v}_2-\Delta\vec{v} 
\end{equation}
and $\vec{\omega}_1'$, $\vec{\omega}_2'$ given by eq. (\ref{drstimp}).
The change in velocity $\Delta\vec{v}$ can be decomposed with
respect to the basis defined above, $\Delta\vec{v} = \gamma_1\vec{u}_1
+ \gamma_2\vec{u}_1^\perp + \alpha\vec{u}_\perp$.  The coefficient
$\alpha$ is given by
\begin{equation}
\alpha = - \left(1 + \frac{ms_{12}^2}{2I}+\frac{ms_{21}^2}{2I}\right)^{-1}
            \vec{V}\vec{u}_\perp,
\end{equation}
while $\gamma_1$ and $\gamma_2$ satisfy the set of linear equations
\begin{equation}
\left(\begin{array}{cc}
A & B \\
B & C
\end{array}\right)
\left(\begin{array}{c}
\gamma_1 \\ \gamma_2
\end{array}\right)
 = -\frac{1+e_t}{2}
\left(\begin{array}{c}
\vec{V}\vec{u}_1 \\ \vec{V}\vec{u}_1^\perp
\end{array}\right)
\end{equation}
with
\begin{equation}
\label{eq:abc}
\begin{split}
A &= 1 + \frac{ms_{21}^2}{2I}(1-(\vec{u}_1\vec{u}_2)^2), \\
B &= -\frac{ms_{21}^2}{2I}(\vec{u}_1\vec{u}_2)
    \sqrt{1-(\vec{u}_1\vec{u}_2)^2}, \\
C &= 1 + \frac{ms_{12}^2}{2I} + \frac{ms_{21}^2}{2I}(\vec{u}_1\vec{u}_2)^2.
\end{split}
\end{equation}

The Liouville operator for two needles must obey the same basic
requirements as for spheres. The only changes are in the condition for
a collision to take place,
\begin{equation}
i \mathcal{T}_+^{(12)} \propto \Theta(L/2-|s_{12}|)\Theta(L/2-|s_{21}|)
                 \delta(|\vec{r}_{12}^\perp|-0^+),
\end{equation}
and in the condition that the two particles are approaching,
\begin{equation}
i \mathcal{T}_+^{(12)} \propto
  \Theta\left(-\frac{d}{dt}|\vec{r}_{12}^\perp|\right).
\end{equation}
Collecting terms  and choosing the correct prefactor gives the
result
\begin{multline}
i \mathcal{T}_+^{(12)} = \left|\frac{d}{dt}|\vec{r}_{12}^\perp|\right|
  \Theta\left(-\frac{d}{dt}|\vec{r}_{12}^\perp|\right) \\ \times
  \Theta(L/2-|s_{12}|)\Theta(L/2-|s_{21}|)\delta(|\vec{r}_{12}^\perp|-0^+) 
  (b_+^{(12)}-1).
\end{multline}
The operator $b_+^{(12)}$ replaces all velocities according to
eqs. \eqref{drstimp} and \eqref{eq:postvelneedle}.

\subsection{Inelastic collision}
The collision rules for rough spheres and needles are easily
generalized to inelastic collisions. This will allow us to set up a
formulation of the dynamics of inelastically colliding grains in terms
of a pseudo-Liouville operator.

\subsubsection{Rough spheres}
Energy dissipation is modeled by normal and tangential restitution.
The collision rules imply for the change in the relative velocity of
the points of contact:
\begin{equation}
\label{eq:rulenormalspherein}
\begin{split}
\hat{\vec{r}}_{12}\vec{V}' &= -e_n\, \hat{\vec{r}}_{12}\vec{V} \\
\hat{\vec{r}}_{12}\times\vec{V}' &= -e_t\, \hat{\vec{r}}_{12}\times\vec{V}.
\end{split}
\end{equation}
The first of these equations describes the reduction of the
normal-velocity component by a non-negative factor $e_n$. This is the
well-known normal restitution. The second equation tries to describe
surface roughness and friction in that it imposes a reduction or even
a reversal of the tangential velocity component. It is motivated by
the picture of small ``bumps'' on the surface which become hooked when
the surfaces are very close.  For all $-1<e_t<+1$ dissipation is
present.

The change in energy is given by
\begin{multline}
\Delta E = -m\Big[\frac{1-e_n^2}{4}(\hat{\vec{r}}_{12}\vec{v}_{12})^2
           + \\ \frac{1-e_t^2}{4}\frac{q}{1+q}
         \Big(\vec{v}_{12}-(\hat{\vec{r}}_{12}\vec{v}_{12})\hat{\vec{r}}_{12}
     -\frac{d}{2}\hat{\vec{r}}_{12}
      \times(\vec{\omega}_1+\vec{\omega}_2)\Big)^2\Big].
\end{multline}
With the parameter range $0\le e_n\le 1$ and $-1\le e_t\le 1$, energy
is only lost and never gained in a single collision.

The conservation laws for linear and angular momenta are unchanged, so
we obtain the same set of equations for the post-collisional velocities
as eqs.  \eqref{eq:postvelsphere}, with however different parameter
values
\begin{equation}
\label{eq:eta}
\eta_n = \frac{1+e_n}{2} \quad\text{and}\quad
\eta_t = \frac{q}{1+q}\frac{e_t+1}{2}.
\end{equation}

Later we will need the inversion of eqs. (\ref{eq:postvelsphere}),
i.e. for given post-collisional velocities we want to determine the
pre-collisional ones.  This is simply done by replacing $e_t$ by
$1/e_t$ and $e_n$ by $1/e_n$ in eqs. (\ref{eq:postvelsphere}).  The
pre-collisional velocities obtained from post-collisional ones will in
the following be denoted by $\vec{v}_1'',\vec{v}_2'',\vec{\omega}_1''$
and $\vec{\omega}_2''$.

\subsubsection{Rough needles}

For hard needles we introduce normal and tangential restitution
according to:
\begin{equation}
\vec{V}'\vec{u}_\perp = -e_n\vec{V}\vec{u}_\perp, \quad
\vec{V}'\vec{u}_1 = -e_t\vec{V}\vec{u}_1, \quad {\rm and} \quad
\vec{V}'\vec{u}_2 = -e_t\vec{V}\vec{u}_2.
\end{equation}
The conservation laws for linear and angular momenta are the same as
for the elastic case, so that one arrives at the same set of eqs.
(\ref{eq:postvelneedle}), the only change affecting the parameter
\begin{equation}
\alpha = -\frac{1+e_n}{2} \left(1 + \frac{ms_{12}^2}{2I}
          +\frac{ms_{21}^2}{2I}\right)^{-1}
            \vec{V}\vec{u}_\perp.
\end{equation}
The energy loss for needles is given by
\begin{multline} \label{needloss}
\Delta E = 
-m\frac{1-e_t^2}{4}\left( \frac{
C(\vec{V}\vec{u}_1)^2 - 2B(\vec{V}\vec{u}_1)(\vec{V}\vec{u}_1^\perp)
+ A(\vec{V}\vec{u}_1^\perp)^2}{AC-B^2}\right) \\
- m\frac{1-e_n^2}{4}\left(1 + \frac{ms_{12}^2}{2I}
          +\frac{ms_{21}^2}{2I}\right)^{-1}(\vec{V}\vec{u}_\perp)^2.
\end{multline}
It can be checked with eqs. \eqref{eq:abc} that the first term is less
than 0 if and only if $-1\le e_t\le 1$. Obviously, the second term is
also less than 0 if $0\le e_n\le 1$. Our method of modeling granular
collisions of needles is therefore consistent with the constraint that
energy may not be gained in a single collision.

\subsection{Time evolution of the distribution function}
\label{adjof}
We will be interested in ensemble averages of observables $f(\Gamma)$
at a time $t$ defined by:
\begin{equation}
\label{eq:average1}
\langle f \rangle (t) = \int d\Gamma\,\rho(\Gamma;0)
                        f(\Gamma;t)=\int d\Gamma\,\rho(\Gamma;t)
                        f(\Gamma).
\end{equation}
Here $\rho(\Gamma;t)$ is the $N$-particle distribution function at
time $t$. The average can either be taken over the \textit{initial}
distribution $\rho(\Gamma;0)$ at time 0, the observable being
propagated to time $t$, or equivalently over the distribution
$\rho(\Gamma;t)$ at time $t$ with the unchanged observable
$f(\Gamma)$. We write eq. (\ref{eq:average1}) as
\begin{equation}
\label{eq:average2}
\langle f \rangle(t) = \int d\Gamma\,\rho(\Gamma;0) e^{i  \mathcal{L} t}
f(\Gamma)
=: \int d\Gamma\, \left( e^{i \overline{\mathcal{L}} t}\rho(\Gamma;0) \right)
   f(\Gamma),
\end{equation}
to define the time-evolution operator $\overline{\mathcal{L}}$ which
describes the time evolution of $\rho$ \cite{explanation}.  To determine
$\overline{\mathcal{L}}$ explicitly, we take the derivative of eq.
\eqref{eq:average2} at time $t=0$ for simplicity,
\begin{equation}
\label{eq:averagediff}
\begin{split}
\partial_t\langle f \rangle(t)\big|_{t=0} &= 
   \int d\Gamma\,\rho(\Gamma;0)i \mathcal{L}
                        f(\Gamma) \\
 &= \int d\Gamma\,
  \left(\partial_t\rho(\Gamma;t)\big|_{t=0}\right)f(\Gamma) =
\int d\Gamma\,\left(i \overline{\mathcal{L}}\rho(\Gamma;0)\right) f(\Gamma).
\end{split}
\end{equation}

The time-evolution operator of the density due to free streaming,
$\overline{{\cal L}}_0$, is easily calculated by partial integration
and we get $ \overline{{\cal L}}_0 = - {\cal L}_0$.  To find an
expression for the time-evolution operator of the density due to
collisions $\overline{{\cal T}}_+^{(12)}$ for spheres, we use eq.
(\ref{eq:averagediff}).  Phase-space coordinates before collision are denoted by
$\Gamma$, after collision by $\Gamma'=b_+^{(12)}\Gamma $ so that
\begin{multline}
  \int d\Gamma \rho(\Gamma;0) i  {\cal T}_+^{(12)}    f(\Gamma) = \\ 
\int d\Gamma \rho(\Gamma;0) \delta (|\vec{r}_{12}|-d)
\Theta(-\frac{d}{dt}|\vec{r}_{12}|) \left|
    \frac{d}{dt}|\vec{r}_{12}| \right|
    \left( f(\Gamma')- f(\Gamma) \right).
\end{multline}
In the first term on the right hand side we make a coordinate
transformation to the variables after collision with Jacobian ${\cal
  J}: =\left|\frac{\partial \Gamma}{\partial \Gamma'}\right|$.  We use
the inverse operator of $b_+^{(12)}$, namely
$b_{-}^{(12)}\Gamma'=\Gamma'' $.  Here the coordinates before
collision in terms of the coordinates after collision are denoted by
$\Gamma''=\Gamma(\Gamma')$.  We note that $\frac{d}{dt}|\vec{r}_{12}|
= \vec{v}_{12} \hat{\vec{r}}_{12}$ and rewrite the first term
\begin{multline}
  \int d\Gamma \rho(\Gamma;0) \delta (|\vec{r}_{12}|-d)
\Theta(-\frac{d}{dt}|\vec{r}_{12}|) \left|
    \frac{d}{dt}|\vec{r}_{12}| \right|
     f(\Gamma') = \\
\int d\Gamma' {\cal J}  \rho(\Gamma'';t) 
\delta (|\vec{r}_{12}|- a)
\Theta(-\vec{v}_{12}'' \hat{r}_{12}) \left|
    \vec{v}_{12}'' \hat{r}_{12} \right|
     f(\Gamma')
\end{multline}
Next we rename $\Gamma'$ by $\Gamma$ and make use of $\vec{v}_{nm}''
\hat{\vec{r}}_{nm} = -\frac{1}{e_n} (\vec{v}_{nm} \hat{\vec{r}}_{nm})
$. This allows us to identify the time-evolution operator of the
distribution function, $\overline{{\cal T}}_+^{(12)}$, by:
\begin{equation}
\label{eq:bol99}
i   \overline{\mathcal{T}}_+^{(12)}= \delta (|\vec{r}_{12}|-d) \left|
    \frac{d}{dt}|\vec{r}_{12}| \right| \left
    (\Theta(\frac{d}{dt}|\vec{r}_{12}|) \frac{{\cal J}}{e_n}b_-^{(12)}
    -\Theta(-\frac{d}{dt}|\vec{r}_{12}|)\right).
\end{equation}

It is common to rewrite eq. (\ref{eq:bol99}) by multiplying it with
$\int d \vec{\sigma} \delta(\vec{\sigma}-\vec{r}_{12})$ so that we can
replace $\vec{r}_{12}$ by $\vec{\sigma}$ in eq. (\ref{eq:bol99}).  In
the second term the integral transformation $\vec{\sigma}\rightarrow
-\vec{\sigma}$ is done and we integrate over $|\vec{\sigma}|$. We
obtain in $D$ dimensions
\begin{equation}
i \overline{{\cal T}}_+^{(12)} = d^{D-1} \int_{\vec{v}_{12}
    \hat{\vec{\sigma}}>0} d \hat{\sigma}\,(\vec{v}_{12}\hat{\vec{\sigma}})\left
    (\frac{{\cal
        J}}{e_n}\delta(\vec{r}_{12}-d\hat{\vec{\sigma}})b_-^{(12)} -
    \delta(\vec{r}_{12}+d\hat{\vec{\sigma}})\right).
\end{equation}
Finally, we note that $t=0$ is not special since we have only chosen
it for the sake of simplicity. Hence we have derived the
time-evolution operator for the $N$-particle distribution function
$\rho(\Gamma;t)$ which is given by the pseudo-Liouville equation
\begin{equation} \label{eq:bol1}
  \partial_t \rho(\Gamma,t) = i \left(- {\cal L}_0(\Gamma) +  \sum_{i<j} 
   \overline{{\cal T}}_+^{(ij)}   \right) \rho(\Gamma,t).
\end{equation}
A similar procedure yields the time evolution operator for the
distribution of needles.

\section{Homogeneous cooling state}
\label{HCS}
We are interested in the time evolution of a gas of freely cooling
rough spheres or needles which is dominated by two-particle
collisions, as discussed in the previous section. We aim at a
description in terms of macroscopic quantities and focus on the decay
in time of the average kinetic energy of translation and rotation,
defined as
\begin{equation}
\label{eq:Ekin}
\begin{split}
\left\langle E_{\rm tr} \right\rangle(t) &= \frac{m}{2N} \sum_i 
 \int d\Gamma \,\rho(\Gamma;t)
\mbox{\boldmath$v$}_i^2 
   =: \frac{D_{\rm tr}}{2} T_{\rm tr}(t)~, \\
\left\langle E_{\rm rot} \right\rangle(t) &= \frac{I}{2N} \sum_i 
\int d\Gamma \,\rho(\Gamma;t)
\mbox{\boldmath$\omega$}_i^2 =: \frac{D_{\rm rot}}{2} T_{\rm rot}(t)~.
\end{split}
\end{equation}
Here $D_{\rm tr}$ and $D_{\rm rot}$ denote the total number of
translational and rotational degrees of freedom respectively.  It is
impossible to compute the above expectation values exactly and we have
to resort to approximations. We assume that the $N$-particle
probability distribution $\rho(\Gamma,t)$ is homogeneous in space and
depends on time only via the average kinetic energy of translation and
rotation:
\begin{equation}
  \label{eq:HCS1}
  \rho_{\rm HCS}(\Gamma;t)_{\rm HCS} \sim
W(\vec{r}_1, \dots ,\vec{r}_N) \,
  \tilde{\rho}
\left(\{\vec{v}_i,\vec{\omega}_i\};T_{\rm tr}(t),T_{\rm rot}(t)\right)~.
\end{equation}
The function $W(\vec{r}_1, \dots ,\vec{r}_N)$ gives zero weight to
overlapping configurations and 1 otherwise. Needles have vanishing volume in
configuration space, so that $W \equiv 1$.  We shall furthermore
assume that $\tilde{\rho}$ factors neglecting correlations of the
velocities of different particles. In the simplest approximation we
take $\tilde{\rho}$  to be  Gaussian in all its momentum
variables
\begin{equation}
\label{eq:HCS4} 
\tilde{\rho}
\left(\{\vec{v}_i,\vec{\omega}_i\};T_{\rm tr}(t),T_{\rm rot}(t)\right)
\propto\exp\left[ - N \left( 
       \frac{E_{\rm tr}}{T_{\rm tr}(t)}+ \frac{E_{\rm rot}}{T_{\rm rot}(t)}
                    \right )\right]~. 
\end{equation}
In the next section, we shall discuss non Gaussian distributions and
shall compute corrections perturbatively.

To determine the time dependence of $T_{\rm tr}(t)$ and $T_{\rm rot}(t)$ we
take time derivatives of eqs. (\ref{eq:Ekin}) and use the
identity $\frac{d}{dt} \langle f \rangle (t) = \int d\Gamma (\frac{d}{dt}
\rho(\Gamma,t)) f(\Gamma) = \int d\Gamma (i \overline{{\cal L}}
\rho(\Gamma,t)) f(\Gamma) = \int d\Gamma \rho (\Gamma,t) i {\cal L}
f(\Gamma)$. Then $\rho(\Gamma,t)$ is replaced by $\rho_{\rm HCS}(\Gamma;t)$,
resulting in
\begin{equation}
\label{eq:HCS6}
\begin{split}
  \frac{d}{dt}  T_{\rm tr}(t)  & 
   = \frac{2}{D_{\rm tr}}
\int d\Gamma \,\rho_{HCS}(\Gamma;t) i {\cal L} {E_{\rm tr}}=
\frac{2}{D_{\rm tr}}
 \left\langle  i {\cal L} {E_{\rm tr}} \right\rangle_{\rm HCS} 
  ~{\rm~~and} \\ 
  \frac{d}{dt}  T_{\rm rot}(t) & 
 =\frac{2}{D_{\rm rot}} 
 \int d\Gamma \,\rho_{HCS}(\Gamma;t) i  {\cal L} {E_{\rm rot}}=
   \frac{2}{D_{\rm rot}} \left\langle i  {\cal L} {E_{\rm rot}} 
                                                      \right\rangle_{\rm HCS}~.
\end{split}
\end{equation}
All that remains to be done are high dimensional phase-space integrals,
the details of which are delegated to appendices \ref{calcsph} and
\ref{calcnee}, for spheres and needles.

\subsection{Results for spheres}
\label{chap:ressph}
After integration over phase space has been performed (see Appendix A
for details), we find
\begin{eqnarray}\nonumber
\frac{D_{\rm tr}}{2} \frac{d}{dt} T_{\rm tr}(t) =
  \left\langle i { \cal L} {E_{\rm tr}} \right\rangle_{\rm HCS} &=& 
  - G A T_{\rm tr}^{3/2} + G B T_{\rm tr}^{1/2} T_{\rm rot} ~,
   \\ 
\frac{D_{\rm rot}}{2} \frac{d}{dt} T_{\rm rot}(t) =
  \left\langle i   { \cal L} {E_{\rm rot}} \right\rangle_{\rm HCS}& =& 
  G B T_{\rm tr}^{3/2} - G C T_{\rm tr}^{1/2} T_{\rm rot} ~, 
  \label{eq:HCS8}
\end{eqnarray}
with the always positive constants $A$, $B$, $C$, and $G$ depending on
space dimensionality $D$.  In two dimensions the constants in eqs. 
(\ref{eq:HCS8}) are given by
\begin{align} \nonumber
G & = 4 d \frac{N}{V}\sqrt{\frac{\pi}{m}} g(d), &\quad
A & =\frac{1-e_n^2}{4}+\frac{\eta_t}{2}(1-\eta_t),            \\ 
B & = \frac{\eta_t^2}{2q},                      &\quad 
C & = \frac{\eta_t}{2q} \left (1-\frac{\eta_t}{q} \right ). 
\label{eq:ressph1}
\end{align}
and in three dimensions
\begin{align} \nonumber
G & = 8 d^2 \frac{N}{V}\sqrt{\frac{\pi}{m}} g(d), &\quad
A & = \frac{1-e_n^2}{4}+\eta_t (1-\eta_t),            \\ 
B & = \frac{\eta_t^2}{q},                      &\quad 
C & = \frac{\eta_t}{q} \left (1-\frac{\eta_t}{q} \right )~. 
\label{eq:ressph2}
\end{align}
The pair correlation function at contact, $g(d)$, is defined in the
usual way \cite{Hansen}.  A detailed discussion of these results, and
in particular the dependence of free cooling on $e_n$ and $e_t$, can
be found in \cite{luding98huth}.

The Enskog value \cite{chapman60,noije97} of the collision frequency
$\omega_E$, i.e. the average number of collisions which a particle
suffers per unit time in $D$ dimensions is given by
\begin{equation}
  \label{eq:ressph3}
\omega_E := 
S_D\frac{N}{V} g(d)  d^{D-1} \sqrt{\frac{T_{\rm tr}(t)}{\pi m}}~. 
\end{equation}
$S_D$ is the surface of a unit sphere in $D$ dimensions. Note that always
$\omega_E \propto G T^{1/2}$.  We define dimensionless time $\tau$ by
$d \tau = \omega_E dt$ so that $\tau$ counts the collisions that on
average each particle has suffered until time $t$.  In a simulation
this would simply be done by counting the number of collisions.  The
functional dependence of the two temperatures on $\tau$ is determined
by
\begin{align}\label{eq:ressph4}
\frac{d}{d \tau} T_{\rm tr} & = - a T_{\rm tr} + b T_{\rm rot}, \\ 
\label{eq:ressph5}
 \frac{d}{d \tau} T_{\rm rot} & = b T_{\rm tr} - c T_{\rm rot} 
\end{align}
with properly defined $a,b,c$. Eq. (\ref{eq:ressph4}) has a simple
interpretation: In a given short interval $\Delta t$ a number of
$\Delta \tau$ collisions occur. Due to these collisions translational
energy decreases by an amount given by the first term, but there is
also a gain term, reflecting that rotational energy is transfered to
translational energy.   The solution of eqs.
(\ref{eq:ressph4},\ref{eq:ressph5}) can be written as
\begin{align}
  T_{\rm tr} &= c_1 K_+ \exp(-\lambda_+ t) + c_2 K_- \exp(-\lambda_-t)~, \\
T_{\rm rot} &= c_1  \exp(-\lambda_+ t) + c_2  \exp(-\lambda_-t)~, \\
K_\pm &= \frac{1}{2b} (c-a \pm \sqrt{(c-a)^2+4b^2})~, \label{eq:ressph7}\\ 
\lambda_\pm &= \frac{1}{2}\left(c+a\mp\sqrt{(c-a)^2+4b^2}\right)~.  
\end{align}
The constants $c_1$ and $c_2$ are determined by the initial conditions
and $\lambda_->0$, $\lambda_+>0$ and $\lambda_->\lambda_+$ holds for
all $e_t$, $e_n$. Hence for long times the ratio of $T_{\rm
  tr}/T_{\rm rot}$ is determined by $K_+$.

\begin{figure}[hbt]
\centering
\makebox[0cm][l]{a)}
\includegraphics[width=.45\textwidth]{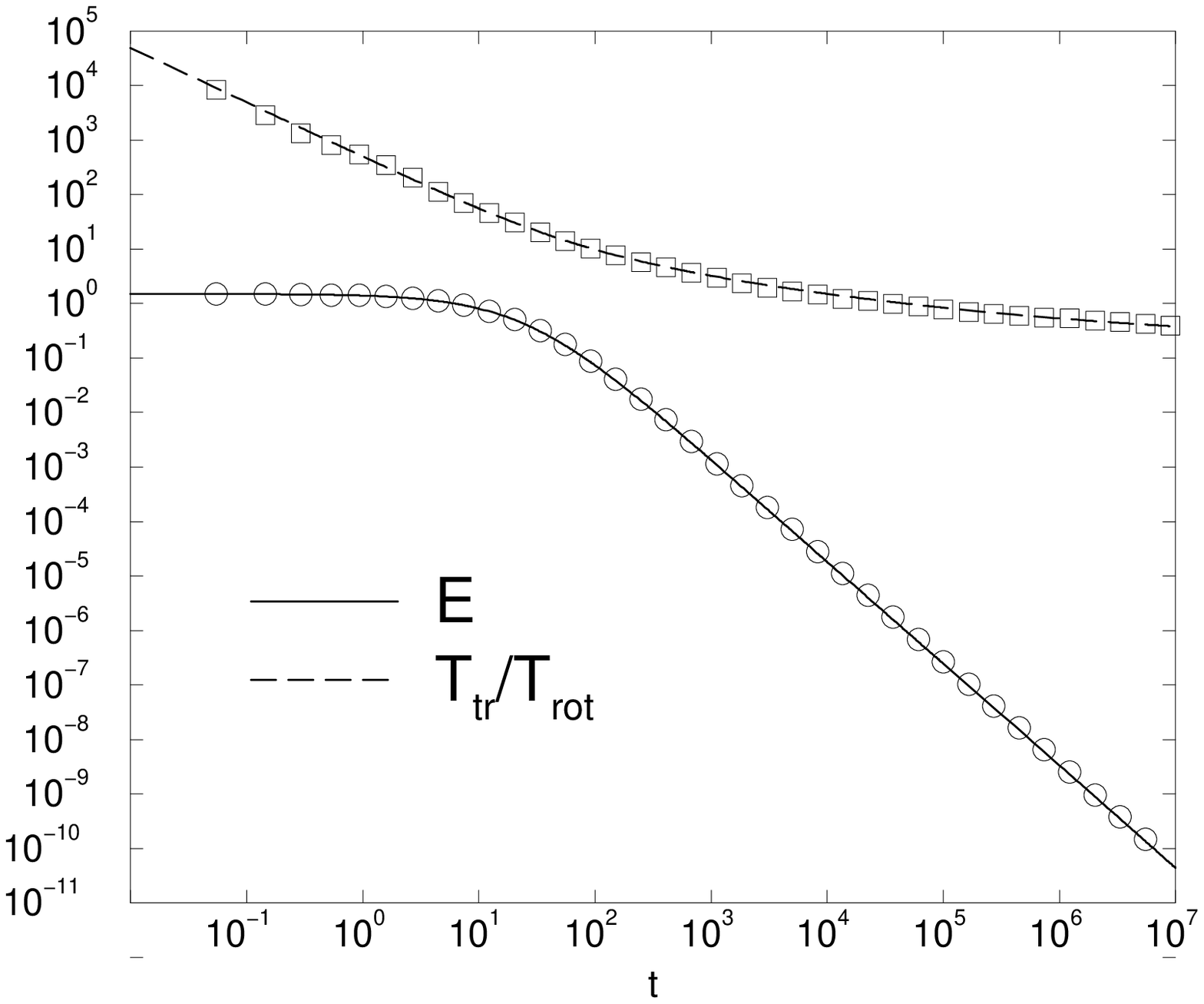}
\hfill
\makebox[0cm][l]{b)}
\includegraphics[width=.5\textwidth]{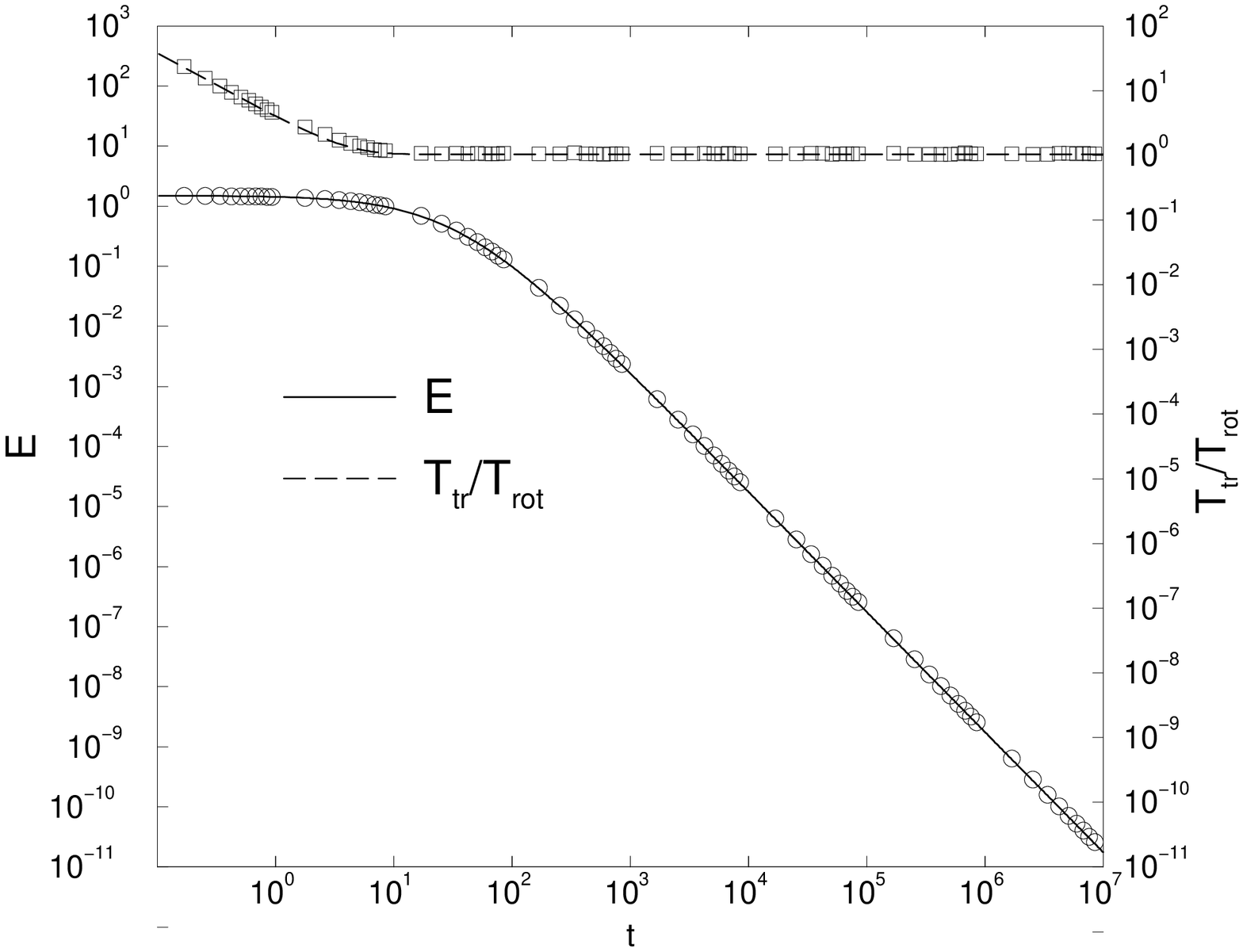}
\caption[]{a) Theoretical prediction (lines) for spheres for the total energy
  $E = \frac{3}{2} (T_{\rm tr}(t)+ T_{\rm rot}(t))$ and the ratio $T_{\rm
    tr}(t)/ T_{\rm rot}(t)$ versus time. Time is plotted in units of
  $\frac{2}{3} G T_{\rm tr}^{1/2}(0)$. We have chosen $e_n=0.9$ and
  $e_t=-0.8$. The symbols represent data of a simulation of 1000 particles
  in a box of length 16 d. \\
  b) The same as a) for needles: The total kinetic energy $E = 3/2
  T_{\text{tr}} + T_{\text{rot}}$ and $T_{\text{tr}}/T_{\text{rot}}$ are
  plotted vs. time in units of $\gamma_n \sqrt{T_{\rm tr}(0)}$. The simulation
  data are from a system of 10000 needles in a box of length 24 $L$ with
  $e_n=0.8$.}
\label{fig:ressph1}
\end{figure}

We now assume that the ratio $T_{\rm tr}/T_{\rm rot} $ has reached its
asymptotic value $K_+$ for some $\tau > \tau_0$ or equivalently
$t>t_0$ and substitute $T_{\rm rot}=T_{\rm tr}/K_+$ into eq.\ 
(\ref{eq:HCS8}) we obtain
\begin{equation}
\frac{d}{d t} T_{\rm tr}  = - F T^{3/2}.
\end{equation}
The resulting equation is of the same functional form as for
homogeneous cooling of smooth spheres, except for the coefficient $F$,
which contains all the dependence on system parameters.  Its solution
is given by
\begin{equation}
  T_{\rm tr} = \frac{T_{\rm tr}(t_0)}{\left[1+T_{\rm tr}
      (t_0)^{1/2}(F/2)(t-t_0)\right]^2} \sim \frac{1}{(Ft/2)^2 }~,
\end{equation}
Haff's \cite{haff83} law of homogeneous cooling.  We have
determined two time scales, first an exponentially fast decay
(measuring time in collisions) towards a state where we find a
constant ratio of translational and rotational energy.  As long as
dissipation is small, we approximate the Enskog-collision frequency for
sufficiently short times by its initial value $\omega(t)\sim
\omega(0)$ so that we find exponential behavior also in real time.
The second stage of relaxation is characterized by a slow,
algebraic decay of both energies, such that their ratio remains
constant. These two time regimes are clearly seen in the numerical
solution of eq. (\ref{eq:ressph1})
for initial conditions $T_{\rm tr}(0)=1$ and $T_{\rm
  rot}(0)=0$, i.e. a system prepared in an equilibrium state of
perfectly smooth spheres. We show in fig. \ref{fig:ressph1} a) in a
double logarithmic plot the time dependence of the total energy $E =
\frac{3}{2} (T_{\rm tr}(t)+ T_{\rm rot}(t))$ and the ratio $T_{\rm
  tr}(t)/ T_{\rm rot}(t)$. Time is plotted in units of $\frac{2}{3} G
T_{\rm tr}^{1/2}(0)$. We have chosen $e_n=0.9$ and $e_t=-0.8$.

\subsection{Results for needles} 
In the case of needles we restrict ourselves to the case of perfectly
smooth needles, i.e. $e_t = -1$. After some lengthy algebra, presented
in appendix \ref{calcnee}, eq. (\ref{eq:HCS6}) can be cast in the
following form
\begin{multline} \label{eq:resnee1}
\frac{2\dot{T}_{\rm tr} }{\gamma_n T_{\rm tr}^{3/2}(1+e_n)} =
-\int_{\Box} d^2r \frac{(1+\frac{T_{\rm rot}}{T_{\rm
        tr}}kr^2)^{1/2}}{1+kr^2} \\
+\frac{1+e_n}{2}
\int_{\Box} d^2r\frac{(1+\frac{T_{\rm rot}}{T_{\rm
        tr}}kr^2)^{3/2}}{(1+kr^2)^2}~,
\end{multline}
\begin{multline}\label{eq:resnee2}
\frac{4\dot{T}_{\rm rot} }{3\gamma_n T_{\rm tr}^{3/2}(1+e_n)} =
-\int_{\Box} d^2r \frac{\frac{T_{\rm rot}}{T_{\rm
      tr}}kr^2 (1+\frac{T_{\rm rot}}{T_{\rm
        tr}}kr^2)^{1/2}}{1+kr^2} \\
+ \frac{1+e_n}{2}
\int_{\Box} d^2r \frac{kr^2 (1+\frac{T_{\rm
        rot}}{T_{\rm tr}} kr^2)^{3/2}}{(1+kr^2)^2}~,
\end{multline}
with $\gamma_n = (2NL^2\sqrt{\pi})/(3V\sqrt{m})$ and $k=(mL^2)/(2I)$.
The two dimensional integration extends over a square of unit length,
centered at the origin.

In fig. \ref{fig:ressph1} b) we plot the numerical solution of eqs.
(\ref{eq:resnee1},\ref{eq:resnee2}) for $e_n = 0.8$ and $k=6$ ($k=6$
corresponds to a homogeneous mass distribution along the rod) as a
function of time in units of $\gamma_n \sqrt{T_{\rm tr}(0)}$. In
addition we have performed simulations of a system of 10000 needles,
confined to a box of length 24 $L$. We show the total kinetic energy
$E = \frac{3}{2} T_{\text{tr}} + T_{\text{rot}}$ (in units of
$T_{\text{tr}}(\tau=0)$) and the ratio $T_{\text{tr}}/T_{\text{rot}}$.
Analytical theory and simulation are found to agree within a few
percent over eight orders of magnitude in time.  ($T_{\rm rot}(0)=0$
has been chosen as initial condition).  For needles we observe an
even clearer separation of time scales.  The decay of
$T_{\text{tr}}/T_{\text{rot}}$ to a constant value $K_+$ happens on a
time scale of order one. In this range of times the total kinetic
energy $E$ remains approximately constant (on a logarithmic scale) and
decays like $t^{-2}$ only {\em after} translational and rotational
energy have reached a constant ratio.  We plug the ansatz $K_+T_{\rm
  rot}=T_{\rm tr}$ into eqs.  (\ref{eq:resnee1},\ref{eq:resnee2}) and
recover Haff's law also for needles. To
determine the constant $K_+$ we use $K_+\dot {T}_{\rm rot}-\dot {T}_{\rm
  tr}=0$, which yields an implicit equation for $c$. Equipartition
holds for {\em all} values of $e_n$ if $k=(mL^2)/(2I)$ is set to
particular value ($k^*=4.3607$), given as the solution of
\begin{equation} \nonumber
 (1-e_n^2) \int_{\Box} d^2r \frac{1-\frac{3}{2} k^*r^2}
  {\sqrt{1+k^*r^2}}=0~.
\end{equation}
For $k<k^*$ we find $ T_{\rm tr} < T_{\rm rot}$ and for $k>k^*$,
$T_{\rm tr} > T_{\rm rot}$. Hence the distribution of mass along the
rods determines the asymptotic ratio of rotational and translational
energy, including equipartition as a special case.

\subsection{Breakdown of homogeneity in dense systems of needles}
It is well known that dense and large systems of inelastically
colliding spheres exhibit clustering so that the assumption of
homogeneity breaks down and deviations from Haff's law of homogeneous
cooling are observed \cite{goldhirsch93,britoernst}. To investigate
inhomogeneities for dense systems of needles we measure hydrodynamic
quantities, i.e we define local variables as the density field, the
translational and rotational flow field and the local rotational and
translational kinetic energy. In order to take local averages over
small volumes, we divide the simulation box into cells whose sizes are
small compared to the box size but large enough to give a reasonable
statistics. We choose cell size, such that on average about 25 needles
are in each cell. For each cell indexed by $\alpha$ we compute the
number density $ \rho_\alpha := \frac{1}{V_{\rm cell}} \sum_{i\in {\rm
    cell}_\alpha } 1 = \langle 1 \rangle_\alpha $, the translational
energy per particle $\rho_\alpha E^{\text{tr}}_\alpha=
\langle\frac{m}{2}\vec{v}_i^2\rangle_{\text{$\alpha$}} $ and the
hydrodynamic temperature $T^{\text{tr}}_\alpha = E^{\text{tr}}_\alpha
- m \vec{U}_\alpha^2/2$ defined by fluctuations around the flow field
$\rho_\alpha \vec{U}_\alpha = \langle\vec{v}\rangle_{\text{$\alpha$}}
$. The corresponding observables of the rotational degrees of freedom
are the rotational energy per particle $E^{\text{rot}}_\alpha$ the
hydrodynamic rotational temperature $T^{\text{rot}}_\alpha$ and the
rotational flow field $\vec{\Omega}_\alpha$.

\begin{figure}[hbt]
\centering
\includegraphics[width=.5\textwidth]{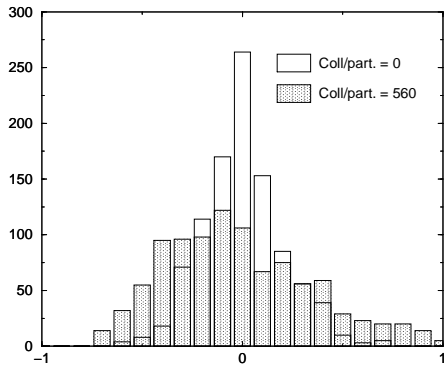}
\caption[]{Histogram of density fluctuations in the initial state and after
  560 Collisions per particle. It is obvious that regions with high
  density have developed.}
\label{fig:brkhom1}
\end{figure}

To check for spatial clustering, we compare the statistics of
fluctuations of the local density, velocity and translational energy
for elastic and inelastic systems. As an example we show in fig.
\ref{fig:brkhom1} the histogram of the deviation of the local density
$\delta \rho_\alpha = \rho_\alpha/n -1 $. We performed simulations of
a dense and large system of 20000 needles confined to a volume with
linear dimension $12 L$ and $e_n=0.9$.
The initial distribution is uniform, corresponding to the equilibrium
state of an elastic system. As the system develops in time with
particles colliding inelastically, we observe that the distribution
broadens, a clear indication that regions of large density have
developed. Histograms of the local translational and rotational
energies look very similar.

\begin{figure}[hbt]
\centering
\includegraphics[width=.8\textwidth]{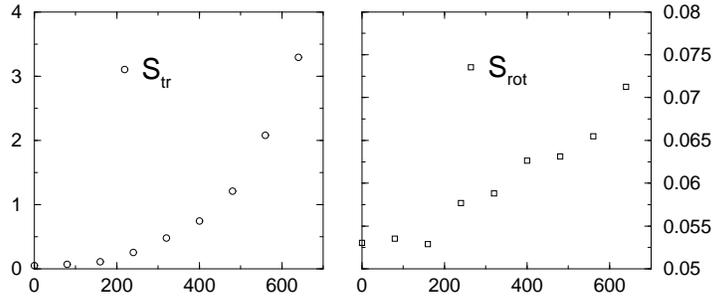}
\caption[]{Ratio $S_{\rm tr}$ ($S_{\rm rot}$) of the local macroscopic
  energy to the local temperature for the 
  translational (rotational) degrees of
  freedom as a function of the number of collisions per particle.}
\label{fig:brkhom2}
\end{figure}

\begin{figure}[hbt]
\centering
\includegraphics[width=.45\textwidth]{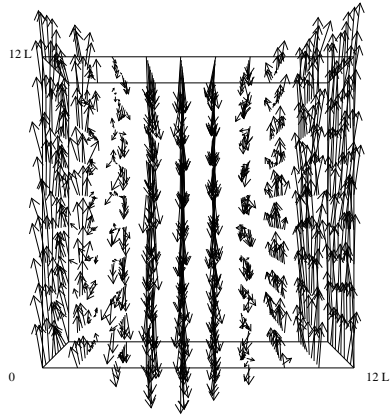}
\caption[]{Flow field after 600 collisions per particle}
\label{fig:brkhom3}
\end{figure}

Inelastic hard spheres without surface roughness tend to move more and
more parallel so that large scale structures in the velocity field
develop. In such a state most of the kinetic energy is to be found in
the energy of the flow field, whereas the energy of the fluctuations
around the flow field is small. A quantitative measure for this effect
\cite{mcnamara96} is the ratio of the total energy of the flow to the
total internal energy of fluctuations: $ S_{\rm tr} := (\sum_\alpha
\frac{m}{2} \rho_\alpha \vec{U}_{\alpha}^2) /(\sum_\alpha \rho_\alpha
T_{\alpha}^{\text{tr}})$ and the analogous quantity $S_{\rm rot}$ for
the rotational degrees of freedom. In fig.  \ref{fig:brkhom2} we show
$S_{\rm tr}$ and $S_{\rm rot}$ as a function of time, measured in
collisions per particle. We observe an increase of $S_{\rm tr}$ by a
factor of 50, whereas $S_{\rm rot}$ increases only by about 50 \%.
Hence the large scale structures in the flow field are much more
pronounced for the translational velocity.
In the fig. \ref{fig:brkhom3} we show the flow field after 600
collisions per particle. We observe two shear bands (note the periodic
boundary conditions) in which the flow field is to a large degree
aligned. In periodic boundary conditions stable shear bands have to be
aligned with the walls of the the box.

\begin{figure}[hbt]
\centering
\includegraphics[width=.6\textwidth]{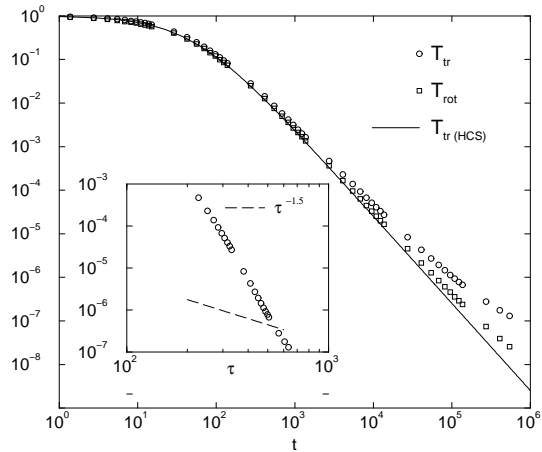}
\caption[]{Data of simulations for  translational and rotational
  temperature as a function of time (units of $\gamma_n \sqrt{T_{\rm
      tr}(0)}$) are compared to the numerical solution of eq.
  (\ref{eq:resnee1},\ref{eq:resnee2}) and to $\tau^{-3/2}$. The inset
  shows $T_{\rm tr}$ as a function of $\tau$ and $\tau^{-3/2}$.  }
\label{fig:brkhom4}
\end{figure}

How does the organization of the flow field influence the decay of the
average energy in the system?  Brito and Ernst \cite{britoernst} have
suggested a generalized Haff's law to describe the time dependence of
the kinetic energy of smooth inelastically colliding spheres even in
the non-homogeneous state.  They found that in the late state where
one finds a well developed flow field the energy decays like
$\tau^{-D/2}$ in $D$ dimensions.  As in section \ref{chap:ressph}
$\tau$ is the average number of collisions suffered by a particle
within time $t$.  In fig.  \ref{fig:brkhom4} we compare the data of
the simulation with the solution of eqs.
(\ref{eq:resnee1},\ref{eq:resnee2}) and in the inset we plot $T_{\rm
  tr}$ as a function of $\tau$ and compare it to $\tau^{-3/2}$. We can
not confirm a $\tau^{-3/2}$ law, but by inspection of fig.
\ref{fig:brkhom3} we see that the range of correlations are already of
the order of the system size, so that finite size effects -- not taken into account in
the theory of Brito and Ernst -- may be dominating. To simulate larger
systems and longer runs has not been possible because simulations of
dense systems are rather time consuming \cite{huthasp}.

\section{Non-Gaussian Distribution}
\label{Expansion}
In this section we keep the assumption of homogeneity and
factorization of the $N$-particle distribution function, but go beyond
the approximation of a purely Gaussian state. Initially the system is
prepared in a Gaussian state, so that deviations from the Gaussian
should be small for short times and perturbation theory can be used to
check the range of validity of the Gaussian approximation. We expand
the one particle distribution function in generalized Laguerre
polynomials (for a definition see \cite{oberhettinger}) around the
Gaussian with time dependent variances. We define an average velocity
$v_0(t) = \sqrt{2 T_{\rm tr}(t)/m}$ and $\omega_0(t) = \sqrt{2 T_{\rm
    rot}(t)/I}$ and scale linear velocities by $v_0(t)$ and angular
velocities by $\omega_0(t)$. The general ansatz for the N-particle
distribution function of the homogeneous cooling state then reads
\begin{multline} \label{eq:full1}
\rho(\Gamma,t)\sim W(\vec{r}_1,...\vec{r}_N)\prod_{i=1}^N 
\rho_i(\vec{v}_i,\vec{\omega}_i,t)~, \quad{\rm and}\\ 
\rho_i(\vec{v}_i,\vec{\omega}_i,t) 
= \frac{1}{Z(t)} 
\exp \left(-\left(\frac{\vec{v}_i}{v_0(t)}\right)^2- 
\left(\frac{\vec{\omega}_i}{\omega_0(t)}\right)^2 \right) \\
\sum_{n,m=0}^{\infty} a_{n,m} (t) L_n^{\alpha} 
\left(\left(\frac{\vec{v}_i}{v_0(t)}\right)^2\right) L_m^{\beta} 
\left(\left(\frac{\vec{\omega}_i}{\omega_0(t)}\right)^2\right)~. 
\end{multline}
We have introduced the abbreviations $\alpha = D_{\rm tr}/2-1 $ and
$\beta = D_{\rm rot}/2-1$.  The average linear and angular velocities,
$v_0(t)$ and $\omega_0(t)$, are time dependent and so are the
coefficients $a_{n,m}(t)$ of the double expansion. At time $t=0$ the
system is equilibrated with temperature $T$ so that $\frac{m}{2}
v_0^2=\frac{D_{\rm tr}}{2}T$ and $\frac{I}{2}
\omega_0^2=\frac{D_{\rm rot}}{2} T$ and hence $a_{n,m}(t)=0$.

The factor $Z(t)$ follows from the proper normalization, 
$\int d \vec{v}_i d \vec{\omega}_i \rho_i = 1$,
\begin{equation}
  \label{eq:full4}
  Z(t)=  v_0^{D_{\rm tr}} \omega_0^{D_{\rm rot}} \sqrt{\pi}^{D_{\rm tr}}
 \sqrt{\pi}^{D_{\rm rot}} a_{0,0}~,
\end{equation}
and we require that $v_0(t)$ and $\omega_0(t)$ be determined by the
conditions
\begin{equation}
  \label{eq:full3}
  \int d \Gamma \vec{v}_1^2 \rho(\Gamma,t) = 
  \frac{D_{\rm tr}}{2} v_0^2(t) \quad {\rm and} 
 \quad \int d \Gamma \vec{\omega}_1^2 \rho(\Gamma,t) = 
\frac{D_{\rm rot}}{2} \omega_0^2(t)~.
\end{equation}
The orthogonality relations of the Laguerre polynomials imply
$\quad a_{1,0}(t)=a_{0,1}(t)=0$ for all times $t$ and
\begin{equation}
  \label{eq:full5}
  a_{n,m}(t) = \frac{1}{\binom{n+\alpha}{n}}
  \frac{1}{\binom{m+\beta}{m}} \int d \Gamma \rho(\Gamma,t) L_n^\alpha
  \left( (\frac{\vec{v}_1}{v_0})^2 \right) L_m^\beta\left(
    (\frac{\vec{\omega}_1}{\omega_0})^2 \right)~.
\end{equation}
The binomial coefficients are denoted by $\binom{a}{b}$ and we choose
$a_{0,0}=1$.

Taking the time derivative of eqs.  (\ref{eq:full3},\ref{eq:full5})
one gets the full time dependence of the homogeneous cooling state
given by the time dependence of all its momenta.
Taking time derivatives of the right hand side of eq. (\ref{eq:full5})
one has to take into account the time dependence of $\rho(\Gamma,t)$, which is
determined by $\overline{{\cal L}}$ as well as the time dependence of
$L_n^\alpha\left( (\frac{\vec{v}_1}{v_0})^2 \right) L_m^\beta\left(
  (\frac{\vec{\omega}_1}{\omega_0})^2 \right)$ via $v_0(t)$ and
$\omega_0(t)$, which follows from eq.(\ref{eq:full3}).

\begin{figure}[hbt]
\centering
\includegraphics[width=.6\textwidth]{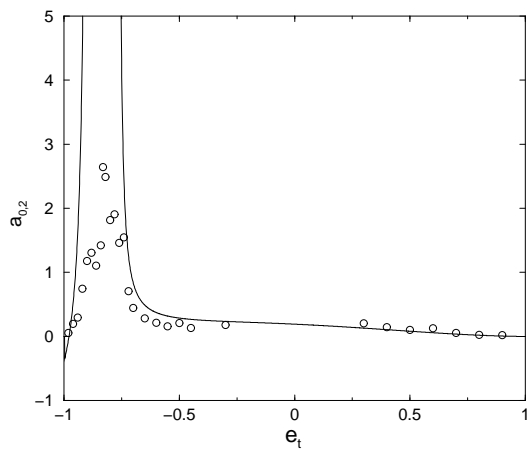}
\caption[]{Coefficient $a_{0,2}$ for $e_n = 0.9$ as a function of
  $e_t$. Theory (straight line) and simulations (circles) are
  compared.}
\label{fig:full1}
\end{figure}

Assuming that all $a_{n,m}$ are stationary in time and that
$v_0/\omega_0=\mu$ is constant we get an infinitely large, nonlinear
system of equations.  To make further progress we truncate the
expansion in eq. (\ref{eq:full1}) and take into account only $a_{n,m}$
for $n+m\leq 2$. We also neglect in the system of equations products
of different $a_{n,m}$, which we assume to be of higher order.  We
show results for $a_{0,2}$ in fig.  \ref{fig:full1} for fixed
$e_n=0.9$ as a function of $e_t$.  Deviations from the Gaussian vanish
for perfectly smooth spheres and are found to increase dramatically
for $e_t \to -0.9$. Deviations from the Gaussian distribution are also
small for perfectly rough spheres which is unexpected, because
rotational degrees of freedom are coupled to translational ones and
$e_n=0.9$. In fact deviations stay small for a broad range of values
of $e_t>\sim -0.75$. We don't consider it meaningful to plot the
theoretical result, once a divergence of $a_{0,2}$ has occurred. We
measured $a_{0,2}$ in simulations of small systems. Thereby we avoid
clustering but have to bear with poor statistics. The simulations
confirm the increase of $a_{0,2}$ around $e_t=-0.7$ in agreement with
the perturbation expansion.

Goldshtein and Shapiro \cite{Goldshtein95} propose a similar set of
momentum equations but they solve it only to lowest order, resulting
therefore in the same asymptotic ratio $\mu$ as given in eq.
(\ref{eq:ressph7}).

\section{Conclusion}
\label{Conclusions}
Two simple models of granular particles with rotational degrees of
freedom are discussed: Rough spheres or discs and, as an example for
non-spherical particles, needles.  We focus on the simplest collision
rules, which allow for a transfer of translational energy to
rotational degrees of freedom. For spheres this is achieved by
tangential restitution (in addition to normal restitution), for
needles normal restitution is sufficient. We show that the time
evolution can be formulated in terms of a pseudo Liouville operator,
thereby generalizing previous work on elastic collisions to inelastic
ones. The presented formalism is general enough to include more
realistic collision rules, for example Coulomb friction for small
angles of impact and tangential restitution for large angles. Work
along those lines can be found in  \cite{herbst99}.

The computation of non-equilibrium expectation values, like e.g. the
relaxation of kinetic energy, require approximations. These are
formulated for the $N$-particle distribution function, which we assume
to be homogeneous in space and depend on time only via the average
translational and rotational energy, $T_{\rm tr}$ and $T_{\rm rot}$.
The distribution of linear and angular momenta is expanded in Laguerre
polynomials around a Gaussian state with time dependent width, $T_{\rm
  tr}$ and $T_{\rm rot}$. The zeroth order approximation, i.e. a pure
Gaussian, leads to two coupled differential equations for the two
temperatures. In both systems, spheres and needles, the relaxation of
translational and rotational kinetic energy is characterized by two
time scales: (1) An exponentially fast decay towards a state with
constant ratio of translational to rotational energy and (2) an
algebraically slow decay of the whole energy, such that the above
ratio keeps constant in time. The theoretically predicted cooling
dynamics is supported by computer simulations of systems of small or
moderate density, where no shearing or cluster instability is observed and the
system remains homogeneous \cite{luding98huth,huthasp}

To study deviations from the Gaussian state, we restrict ourselves to
rough spheres and truncate the expansion in Laguerre polynomials,
keeping the first three terms (the first order term for smooth spheres
has already been computed in \cite{twanexpan}). This perturbative
approach is shown to break down for certain values of $e_t$ and $e_n$,
where deviations from the Gaussian are shown to diverge. These results
are confirmed by simulations.  We indicate, how a more general
expansion with time dependent coefficients can be achieved. For
totally smooth spheres this expansion has been performed up to fifth 
order \cite{huthmannbrito}. It predicts an exponentially fast decay of
the coefficients to their stationary values in agreement with
simulations and direct solutions of the Boltzmann equation
\cite{brey}.

For needles we observe and investigate the breakdown of homogeneity in
simulations of dense systems, where the inter-particle spacing is
smaller than the length of the needles. Large-scale structures in the
translational velocity field are seen to develop. Furthermore the
density does not remain homogeneous but clusters form and dissolve
again. These effects lead to deviations from the solution of the
homogeneous cooling state on the longest times scales and a third
stage of cooling is found. It is characterized by an even slower decay
of the kinetic energy, most of the energy being stored in the
macroscopic velocity field.

We plan to derive generalized hydrodynamic equations for grains with
rotational degrees of freedom and in particular hard rods.  Such a set
of hydrodynamic equations could serve as a starting point for a
stability analysis, similar to the work of Brito and Ernst
\cite{britoernst} for smooth spheres.

{\em Acknowledgement.}-- This work has been supported by the DFG through
SFB 345 and Grants Zi209/5-1 and Zi209/6-1.

\begin{appendix}
\section{Calculations for spheres}
\label{calcsph}
In this appendix we explain, as an example, the main steps to
calculate $\left\langle i {\cal L} E_{\rm tr}\right\rangle_{\rm HCS}$
of eq. (\ref{eq:HCS6}) in 2D.  We define the configuration integral
\begin{equation}
  \label{eq:app1.1}
  Q^N := \int \prod_{i=1}^N d \vec{r}_i W(\vec{r}_1,\ldots,\vec{r}_N) ~.
\end{equation}
The proper normalized $N$-particle distribution function for the
HCS-state reads
\begin{multline}
  \rho_{\rm HCS}(\Gamma;t) = \frac{1}{Q^N}
  W(\vec{r}_1,\ldots,\vec{r}_N) \left(\frac{m}{2\pi T_{\rm tr}(t)}
  \right)^{N}
  \left(\frac{I}{2\pi T_{\rm rot}(t)}\right)^{N/2} \times \\
  \exp\left[-\sum_{i=1}^N \left( \frac{m}{2 T_{\rm tr}(t)}
      \vec{v}_i^2+ \frac{I}{2T_{\rm rot}(t)}\omega_i^2\right)
  \right]~. \label{eq:app1.2}
\end{multline}
The angular velocity is a scalar in two dimensions, but a vector in
more than two dimensions.  Free streaming does not change the energy,
so we have to take into account only the collision operator ${\cal
  L}_+$ and compute
\begin{multline}
\left\langle i {\cal L}_+ E_{\rm tr}\right\rangle_{\rm HCS} = 
\frac{1}{2}\sum_{i\ne j}
\int d\Gamma \,
\rho_{\rm HCS}(\Gamma;t) 
i {\cal T}_+^{(ij)} \frac{1}{N} \sum_{k=1}^{N} \frac{m}{2} \vec{v}_k^2=\\
\frac{1}{2N}\sum_{i\ne j}
\int d\Gamma \,
\rho_{\rm HCS}(\Gamma;t) 
i {\cal T}_+^{(ij)} \frac{m}{2}
\left(\vec{v}_i^2+\vec{v}_j^2\right)~. 
\label{eq:app1.3}
\end{multline}
The binary collision operator ${\cal T}_+^{(ij)}$ gives a contribution
only, if either $k=i$ or if $k=j$. Next, we introduce two
$\delta$-functions,
\begin{multline}
\left\langle  i {\cal L}_+ E_{\rm tr}\right\rangle_{\rm HCS} =            
\frac{1}{2N}\sum_{i\ne j}
\int d \Gamma
\int d \vec{R}_1 d\vec{R}_2 
\delta(\vec{R}_1-\vec{r}_i)
\delta(\vec{R}_2-\vec{r}_j)          \\
\rho_{\rm HCS}(\Gamma;t) 
i {\cal T}_+^{(ij)} \frac{m}{2} 
\left(\vec{v}_i^2+\vec{v}_j^2\right)~, 
\label{eq:app1.4}
\end{multline} 
which allows us to replace $\vec{r}_i$ by $\vec{R}_1$ and $\vec{r}_j$
by $\vec{R}_2$ in ${\cal T}_+^{(ij)}$.  We define the pair correlation
function $g(|\vec{R}_1-\vec{R}_2|)$ by
\begin{multline}
  \label{eq:app1.5}
  \frac{N}{V^2} g(|\vec{R}_1-\vec{R}_2|) := \\ \frac{1}{N} \sum_{i\ne j}
  \frac{1}{Q^N} \int \prod_{k=1}^N d\vec{r}_k
  W(\vec{r}_1,\ldots,\vec{r}_N)
  \delta(\vec{R}_1-\vec{r}_i)\delta(\vec{R}_2-\vec{r}_j)\;.
\end{multline}
Eq. (\ref{eq:app1.5}) is used to rewrite eq. (\ref{eq:app1.4}) in
terms of the pair correlation function.  Integration over all
velocities and angular velocities with index $k$ and $i\ne k \ne j$
gives $1$ due to normalization. We get
\begin{multline}
 \left\langle  i {\cal L}_+ E_{\rm tr} \right\rangle_{\rm HCS} = 
   \frac{N}{2 V^2} 
  \left(\frac{m}{2\pi  T_{\rm tr}(t)}\right)^{2}
  \frac{I}{2\pi T_{\rm rot}(t)} 
  \int d \omega_1 d\omega_2
  d \vec{R}_1 d\vec{R}_2 
  d \vec{v}_1 d\vec{v}_2                  \\
  \exp\left(
    -\frac{m}{2 T_{\rm tr}(t)} (\vec{v}_1^2+\vec{v}_2^2) -
    \frac{I}{2T_{\rm rot}(t)}(\omega_1^2+\omega_2^2 )  \right)            \\
  g(\vec{r}) 
  \left|\vec{v}_{12}\cdot\hat{\vec{r}}\right|
  \Theta\left(- \vec{v}_{12}\cdot\hat{\vec{r}}\right)
  \delta\left(|\vec{r}|-d \right) \Delta E_{\rm tr}~.
\end{multline}
The loss of translational energy of two colliding particles is denoted
by $\Delta E_{\rm tr}$.  We use the abbreviation
$\vec{R}_1-\vec{R}_2=\vec{r}=r\hat{\vec{r}}$ and neglect non
contributing terms linear in $\Omega$ so that $\Delta E_{\rm tr}$ is
given by
\begin{multline}
  \Delta E_{\rm tr}=  \frac{m}{2}
  \big[2\eta_t(\eta_t-1)(\vec{v}_{12}^2-(\vec{v}_{12}
  \cdot\hat{\vec{r}})^2)-       \\ 
  (1/2)(1-e_n^2)(\vec{v}_{12}\cdot\hat{\vec{r}})^2
  + (1/2) \eta_t^2 d^2 (\omega_1+\omega_2)^2 \big]~.
\end{multline}
To perform the remaining integrations we substitute
\begin{alignat}{2}
\Omega & =\frac{1}{\sqrt{2}}(\omega_1+\omega_2),
\quad &  \omega & =\frac{1}{\sqrt{2}}(\omega_1-\omega_2), \\
\vec{V} & =\frac{1}{\sqrt{2}}(\vec{v}_1+\vec{v}_2),
\quad & \vec{v} & =\frac{1}{\sqrt{2}}(\vec{v}_1-\vec{v}_2),\\
\vec{r} & =\vec{R}_1-\vec{R}_2, \quad & \vec{R} & =\vec{R}_1.
\end{alignat}
The Jacobian determinant for the above transformation is $1$.
Integration over $\omega$, $\vec{V}$ and $\vec{R}$ all give the value
$1$ due to normalization.  We are left with
\begin{multline} \nonumber
  \left\langle i {\cal L}_+ E_{\rm tr}\right\rangle_{\rm HCS} = 
 \frac{N}{V}  \frac{m}{2\pi T_{\rm tr}(t)} 
  \left(\frac{2I}{2\pi T_{\rm rot}(t)}\right)^{1/2}  
  \int d \Omega d \vec{r} d\vec{v} \\
  \exp \left(
    -\frac{m\vec{v}^2}{2
      T_{\rm tr}(t)} - \frac{I\Omega^2}{2T_{\rm rot}(t)}
  \right)   
  g(\vec{r}) \left|\vec{v}\cdot\hat{\vec{r}}\right|
  \Theta \left(
    -\vec{v}\cdot\hat{\vec{r}}
  \right)
  \delta\left(|\vec{r}|-d \right)\\
 \frac{m}{2} \big [
  2\eta_t(\eta_t-1)(\vec{v}^2-
  (\vec{v}\cdot\hat{\vec{r}})^2)
  -(1/2)(1-e_n^2)(\vec{v}\cdot\hat{\vec{r}})^2
  +(1/2)\eta_t^2 d^2 \Omega^2 
  \big ]~.
\end{multline}
The integration over $|\vec{r}|$ yields $dg(d)$. Choosing e.g.
$\vec{r}$ to point along the $x$-axis, the integrals over linear and
angular velocities can easily be done as moments of a Gaussian
distribution. The result is independent of $\hat{\vec{r}}$, so that
the integration over $\hat{\vec{r}}$ gives $2\pi$.  Finally we obtain
the result of eq. (\ref{eq:ressph1}).

\section{Calculations for needles}
\label{calcnee}
In this appendix we present some of the detailed calculations for
needles. As a first step we express the orientation of the rods in
spherical coordinates
$\vec{u}_i=(\sin(\theta_i)\cos(\phi_i),\sin(\theta_i)\sin(\phi_i),
\cos(\theta_i))$. The canonical momenta (translational and rotational)
are then given by
\begin{equation}
\vec{p}_i = m \vec{v}_i~, \qquad  p_{\theta_i}= I \dot{\theta_i}~,
  \qquad p_{\phi_i} =I\dot{\phi_i} 
\sin^2\theta~. \label{eq:app2.1}
\end{equation}
In the following calculation it will be necessary to express
$\dot{\vec{u}}_i$ in terms of canonical momenta
\begin{equation}
\dot{\vec{u}}_i =
\frac{p_{\theta_i}}{I} \vec{e}_{\theta_i} +   \frac{p_{\phi_i}}{\sin 
  \theta_i I} \vec{e}_{\phi_i}~.  
\label{eq:app2.2}
\end{equation}
$\vec{e}_{\theta_i}$ and $\vec{e}_{\phi_i}$ are orthogonal unit vectors
in $\theta_i$ and $\phi_i$ direction.
The kinetic energies per particle  are  then given by
\begin{equation}\label{eq:app2.3}
  E_{\rm tr} 
    = \frac{1}{N} \sum_{i=1}^{N}
  \frac{1}{2 m}\vec{p}_i^2, \quad E_{\rm rot}= \frac{1}{N}\sum_{i=1}^{N} \frac{1}{2I}p_{\theta_i}^2+
  \frac{1}{2I\sin^2\theta_i} p_{\phi_i}^2~. 
\end{equation}
We want to calculate non-equilibrium expectation values with the
normalized probability distribution given in eq. (\ref{eq:HCS4}).  We
consider again as an example the translation energy per particle
$E_{\rm tr}$.
\begin{multline}
\left\langle i {\cal L_+} E_{\rm tr} \right\rangle = 
  \frac{1}{V^N}\frac{1}{(4\pi)^N}\frac{1}{(2\pi M T_{\rm
      trans})^{3N/2}} \frac{1}{(2 \pi I T_{\rm rot})^{N}} \\
  \frac{1}{2} \sum_{m \ne n} \int \prod_{j=1}^{N} d\vec{r}_j\;
  d\phi_j\; d\theta_j\; d \vec{p}_j\; d p_{\theta_j}\; dp_{\phi_j} \\
  \exp[- N E_{\rm tr}/T_{\rm tr}(t)-N  E_{\rm
      rot}/T_{\rm rot}(t)] i {\cal T}_+^{(nm)} E_{\rm tr}~.
  \label{eq:app2.4}
\end{multline}
Similar to the calculation for the spheres we see that the binary
collision operator ${\cal T}_+^{(nm)}$ gives a contribution only, if
either $i=n$ or if $i=m$. We can sum over $N(N-1)$ identical integrals
and get
\begin{multline}
  \frac{N-1}{2 V^2}\frac{1}{(4\pi)^2} \frac{1}{(2\pi m T_{\rm tr})^3}
  \frac{1}{(2 \pi I T{\rm rot})^2} \int \prod_{j=1}^{2} d\vec{r}_j\;
  d\phi_j\; d\theta_j\; d \vec{p}_j\; d p_{\theta_j}\; dp_{\phi_j} \\ 
  \exp[-E_{\rm tr}^{12}/T_{\rm tr}(t)- E_{\rm rot}^{12}/T_{\rm rot}(t)]
  \left|\frac{d}{dt}\left|\vec{r}_{12}^\perp\right|\right| \Theta(-\frac{d}{dt}\left|\vec{r}_{12}^\perp\right|)\\
  \Theta(L/2-|s_{12}|) \Theta(L/2-|s_{21}|)
 \delta(|\vec{r}_{12}^\perp|-0^+) 
  \Delta E_{\rm tr}^{12}~. \label{eq:app2.5}
\end{multline}
$E_{\rm tr}^{12}$ ($E_{\rm rot}^{12}$) is the sum of the translational
(rotational) kinetic energy of particle 1 and 2 and with $\Delta
E_{\rm tr}^{12}$ we denote the change of the translational kinetic
energy of particle 1 and 2 in a collision:
\begin{align}
  \Delta E_{\rm tr}^{12} & = \frac{(\vec{p}_{1}-\vec{p}_2)\cdot \Delta\vec{p}}{m}+
  \frac{\Delta\vec{p}^2}{m}~, \\
\Delta\vec{p} &= - \frac{1+e_n}{2} \frac{1}{\frac{1}{m} + 
 \frac{s_{12}^2}{2I} + \frac{s_{21}^2}{2I}}(\vec{V} \cdot \vec{u}_\perp) \vec{u}_\perp ~.
\end{align}
$\vec{V}$ is the relative velocity of the contact points defined in
eq. (\ref{vrel}).

We introduce relative coordinates $\vec{r}_{12} =
\vec{r}_1-\vec{r}_{2}$ and $\vec{r}=\vec{r}_1$ and the variables
\begin{align} \nonumber
  z &:= \vec{r}_{12} \cdot  \vec{u}_\perp ~,\\ \nonumber
  a &:= \vec{r}_{12} \cdot \vec{u}_1 - \frac{\vec{u}_1\cdot\vec{u}_2}
{\sqrt{1-(\vec{u}_1\cdot\vec{u}_2)^2}}
  \vec{r}_{12}\cdot\vec{u}_1^\perp = -s_{12} ~,\\ \nonumber
  b &:= \frac{1}
{\sqrt{1-(\vec{u}_1\cdot\vec{u}_2)^2}}\vec{r}_{12}\cdot\vec{u}_1^\perp
=s_{21}~.
\end{align}
The Jacobian of the transformation is given by
$\sqrt{1-(\vec{u}_1\cdot\vec{u}_2)^2}$.  We remark that
$\frac{d}{dt}\left|\vec{r}_{12}^\perp\right|=
\vec{V}\cdot\vec{u}_\perp {\rm sign}(\vec{r}_{12}\cdot\vec{u}^\perp)$
and we find again the relative velocity of the contact points $\vec{V}
= \frac{\vec{p}_{12}}{m}-a \dot{\vec{u}}_1 -b\dot{\vec{u}}_2$ given in
the new coordinates. Integration over $\vec{r}$ gives $V$ and
integration over $z$ gives the sum of two $\Theta$--functions
$\Theta(\pm \vec{V}\cdot\vec{u}_\perp)$. This reflects the fact that if one
particle touches the other from 'above' the sign of the relative
velocity of the contact point has to be negative, if the particle
touches from 'below' the velocity  has to be positive. Next one introduces
relative and center of mass momenta as well as dimensionless
variables:
\begin{align}\nonumber
\vec{\chi} & :=  \frac{1}{\sqrt{2 m T_{\rm
      trans}}}(\vec{p}_1-\vec{p}_2)~, & \quad 
\vec{\gamma} & : = \frac{1}{\sqrt{2 m T_{\rm
      trans}}}(\vec{p}_1+\vec{p}_2)~, \\ \nonumber
\tilde{p}_{\theta_i} & := \frac{p_{\theta_i}}{\sqrt{I T_{\rm rot}}}~,
& \quad 
\tilde{p}_{\phi_i} & :=  \frac{p_{\phi_i}}{\sqrt{I T_{\rm rot}} \sin\theta_i}.
\end{align}
The integration over $\vec{\gamma}$ can be done and the result is
proportional to
\begin{multline}
\sum_{p =\pm 1}
\int da \; db \; d\phi_1 \sin\theta_1d\theta_1\; d\phi_2
\sin\theta_2 d\theta_2   \; d\vec{\chi} \; d\tilde{p}_{\theta_1}\; d\tilde{p}_{\phi_1}
d\tilde{p}_{\theta_2}\; d\tilde{p}_{\phi_2}
\\ \sqrt{1-(\vec{u}_1\cdot \vec{u}_2)^2}
\exp[-\frac{1}{2}(\vec{\chi}^2+\tilde{p}_{\phi_1}^2+\tilde{p}_{\phi_2}^2+\tilde{p}_{\theta_1}^2+\tilde{p}_{\theta_2}^2
)] \\
\left|\tilde{\vec{V}} \cdot \vec{u}_\perp \right| \Theta\left(p \left|
    \tilde{\vec{V}} \cdot \vec{u}_\perp \right|\right) 
\Theta(|a|-L/2)\Theta(|b|-L/2)
\Delta E_{12}~, \label{eq:app2.7}
\end{multline}
all expressed in new variables and $\dot{\vec{u}}$ by eq.
(\ref{eq:app2.2}), e.g.
\begin{multline}
\tilde{\vec{V}} = \sqrt{2 T_{\rm trans}/m} \vec{\chi} - a \sqrt{T_{\rm
 rot}/I} (\tilde{p}_{\theta_1}\vec{e}_{\theta_1} + \tilde{p}_{\phi_1}\vec{e}_{\phi_1})-\\
  b \sqrt{T_{\rm
 rot}/I} (\tilde{p}_{\theta_2}\vec{e}_{\theta_2} +
\tilde{p}_{\phi_2}\vec{e}_{\phi_2})~. \label{eq:app2.8}
\end{multline}
We want to perform the remaining Gaussian integrals, but we have
expressed different terms either in $(\vec{u}_i^\perp,\vec{u}^\perp)$
defined according to eq. (\ref{orthosys}) with $i=1,2$ or in
$(\vec{e}_{\theta_i},\vec{e}_{\phi_i})$.  It is useful to note that
$(\vec{u}_i^\perp,\vec{u}^\perp)$ and
$(\vec{e}_{\theta_i},\vec{e}_{\phi_i})$ are {\em two different}
orthonormal basis of the plane perpendicular to $\vec{u}_i$, so that
we can make a orthogonal coordinate transformation from one system to
the other. The variables $\tilde{p}_{\theta_i}$ and
$\tilde{p}_{\phi_1}$ are now standard normally distributed and after a
orthogonal coordinate transformation the new coordinates will again be
standard normally distributed. This means we can equivalently write
$(\tilde{p}_{\theta_i}\vec{e}_{\theta_i} +
\tilde{p}_{\phi_i}\vec{e}_{\phi_i})$ as $ (v_i \vec{u}_i^\perp + w_i
\vec{u}^\perp) $ with standard normally distributed variables $v_i$
and $w_i$. With this definition of $v_i$ and $w_i$ we are able to
evaluate for example terms of the form $
(\tilde{p}_{\theta_1}\vec{e}_{\theta_1} +
\tilde{p}_{\phi_1}\vec{e}_{\phi_1})\cdot\vec{u}_\perp \equiv ( v_1
\vec{u}_1^\perp + w_1 \vec{u}_\perp)\cdot \vec{u}_\perp = w_1 $, where
we used $\vec{u}_1^\perp \cdot \vec{u}^\perp = \vec{u}_2^\perp \cdot
\vec{u}^\perp=0 $.  We can integrate freely over $v_1$ and $v_2$ and
the two components of $\chi$ perpendicular to $\vec{u}_\perp$. We
denote with $d\Omega_i:= d\phi_i\sin(\theta_i)d\theta_i$ and the
intermediate result reads
\begin{multline}
\sum_{p=\pm 1}
  \frac{N-1}{2V} \frac{1}{(4\pi)^2} \frac{1}{(2\pi)^{(3/2)}} \int
  da~db~d\vec{s}~ d\Omega_1 d\Omega_2 \exp(-\frac{1}{2} \vec{s}^2)
  \sqrt{1-(\vec{u}_1\cdot\vec{u}_2)^2} \\
  |\vec{G}\cdot\vec{s}| \Theta(p\vec{G}\cdot\vec{s})\Big[ -s_1
  \sqrt{\frac{2 T_{\rm tr}}{m} }\frac{1+e_n}{2}
  \frac{1}{\frac{1}{m}+ \frac{a^2}{2I} + \frac{b^2}{2I} }
  \vec{G}\cdot\vec{s}+ \\ \frac{1}{m} \left(\frac{1+e_n}{2}\right)^2
 \left(\frac{1}{\frac{1}{m}+ \frac{a^2}{2I} + \frac{b^2}{2I}}\right)^2
   (\vec{G}\cdot\vec{s})^2   \Big]~.
\end{multline}
We introduced the vectors
$\vec{s}:=(s_1,s_2,s_3):=(\vec{\chi}\cdot\vec{u}_\perp,w_1,w_2)$ and
$\vec{G} = \left(\sqrt{\frac{2 T_{\rm tr}}{m}}, -a \sqrt{\frac{T_{\rm
        rot}}{I}}, - b\sqrt{\frac{T_{\rm rot}}{I}}\right)$.

We can now perform the integral over $\vec{s}$. We sketch here only
how this is done. We want to integrate
\begin{equation}
  \int d\vec{s}~
  \exp(-\frac{1}{2}\vec{s}^2)\Theta(\pm \vec{G}\cdot\vec{s})
  |\vec{G}\cdot\vec{s}| (\vec{G}\cdot\vec{s}) s_1~. \label{eq:app2.100}
\end{equation}
Let $(\vec{e}_1,\vec{e}_2,\vec{e}_3)$ be the original coordinate
system and we define a coordinate system
$(\vec{e}_x,\vec{e}_y,\vec{e}_z)$ in which the $z$-axis is parallel to
$\vec{G}$ and we decompose $\vec{s}$ in this coordinate system
$\vec{s}=( s_x,s_y,s_z)$.  Then eq. (\ref{eq:app2.100}) reads
\begin{multline}
  \int d s_x d s_y d s_z \exp(-\frac{1}{2}(s_x^2+s_y^2+s_z^2))
  \Theta(\pm s_z) \\
  |\vec{G}||s_z||\vec{G}|s_z   
 \left[ (s_x\vec{e}_x+s_y\vec{e}_y
  +s_z\vec{e}_z)\cdot\vec{e}_1\right]~.
\end{multline}
Only the term, which is proportional to $s_z\vec{e}_z$, contributes
and the Gaussian integral can easily be performed. Using that
$|\vec{G}|\vec{e}_z = \vec{G}$ we write
$|\vec{G}|\vec{e}_z\cdot\vec{e}_1 = \vec{G}\cdot \vec{e}_1= G_1 $ and
we end up with the result $ 4 \pi |\vec{G}| G_1 $. Only the integrals
over $\Omega_1$ and $\Omega_2$ have to be done with standard
techniques.  All other integrals are performed similarly and the
results are quoted in the main text.

\end{appendix}

%INDEX%%%%%%%%%%%%%%%%%%%%%%%%%%%%%%%%%%%%%%%%%%%%%%%%%%%%%%%%%%%%%%%
\clearpage
\addcontentsline{toc}{section}{Index}
\flushbottom
\printindex
%%%%%%%%%%%%%%%%%%%%%%%%%%%%%%%%%%%%%%%%%%%%%%%%%%%%%%%%%%%%%%%%%%%%%


\begin{thebibliography}{99}
\bibitem{explanation} It was shown in \cite{ernst69} and
  \cite{noije97} that the time evolution operator can be represented
  in the form exp($i \overline{\mathcal{L}} t$) without generating
  overlap configurations.
\end{thebibliography}
\end{document}